\documentclass[12pt]{article}
\usepackage{graphicx,amsmath}
\usepackage{units}
\usepackage{color}
\usepackage[utf8]{inputenc}

\newlength{\dinwidth}
\newlength{\dinmargin}
\def\lapproxeq{\lower .7ex\hbox{$\;\stackrel{\textstyle                                                    
<}{\sim}\;$}}                                                    
\def\gapproxeq{\lower .7ex\hbox{$\;\stackrel{\textstyle                                                    
>}{\sim}\;$}}                                                    
\def\be{\begin{equation}}                                                    
\def\ee{\end{equation}}                                                    
\def\bea{\begin{eqnarray}}                      
\def\eea{\end{eqnarray}}

\def\bb{\vec{b}'}

\def\bb{b\bar{b}}
\def\cc{c\bar{c}}

\def\GeV{\rm GeV}
\def\TeV{\rm TeV}

\def\sh{\hat s}
\def\sh2{{\hat s}^2}

\def\MS{\overline{\rm MS}}

\begin{document}

\begin{flushright}                                                    
LCTS/2015-30  \\
IPPP/15/60  \\
DCPT/15/120 \\                                                    
\today \\                                                    
\end{flushright} 

\vspace*{0.5cm}

\begin{center}
{\Large \bf Charm and beauty quark masses in the MMHT2014}\\ 
\vspace*{0.5cm}{\Large \bf global PDF analysis}\\

\vspace*{1cm}
L. A. Harland-Lang$^{a}$, A. D. Martin$^b$, P. Motylinski$^a$ and R.S. Thorne$^a$\\                                               
\vspace*{0.5cm}                                                    
                                                  
$^a$ Department of Physics and Astronomy, University College London, WC1E 6BT, UK \\           
$^b$ Institute for Particle Physics Phenomenology, Durham University, DH1 3LE, UK                                                    
                                                    
\vspace*{1cm}

\begin{abstract}
\noindent We investigate the variation in the MMHT2014 PDFs when we 
allow the heavy quark masses $m_c$ and $m_b$ to vary away from their 
default values. We make PDF sets available in steps of 
$\Delta m_c =0.05~\GeV$ and $\Delta m_b =0.25~\GeV$, and present the 
variation in the PDFs and in the predictions. We examine the  
comparison to the HERA data on charm and beauty structure functions 
and note that in each case the heavy quark data, and the inclusive
data, have a slight preference for lower masses than our default values.
We provide PDF sets with 3 and 4 active quark flavours, as well as the 
standard value of 5 flavours.  
We use the pole mass definition of the quark masses, as in the default 
MMHT2014 analysis, but briefly comment on the $\MS$ definition. 

\end{abstract}
                                                   
\vspace*{0.5cm}                                                    
                                                    
\end{center}

\section{Introduction  \label{sec:1}} 

Over the past few years there has been a significant improvement in both the precision and in the variety of 
the data for deep-inelastic and related hard-scattering processes. Since the MSTW2008 analysis~\cite{MSTW} we have seen the appearance of the 
HERA {\it combined} H1 and ZEUS data on the total~\cite{H1+ZEUS} and also on the charm structure functions~\cite{H1+ZEUScharm}, together with a 
variety of new hadron-collider data sets from the LHC, and in the form of 
updated Tevatron 
data (for full references see~\cite{MMHT}). Additionally, the procedures used 
in the global PDF analyses of data have 
been improved, allowing the parton distributions of the proton to be determined 
with more precision and with more confidence. This allows us to improve 
predictions for Standard Model signals and to model Standard Model backgrounds
to possible experimental signals of New Physics 
more accurately.  One 
area that now needs careful attention, at the present level of accuracy, is the 
treatment of the masses of the charm and beauty quarks, $m_c$ and $m_b$, in 
the global analyses.  
Here we extend the recent MMHT2014 global PDF analysis~\cite{MMHT} to study the 
dependence of the PDFs, and the quality of the comparison to data, under 
variations of these masses away from their default values of $m_c=1.4~\GeV$ and 
$m_b=4.75~\GeV$, as well as the resulting predictions for processes at the LHC. We make 
available central PDF sets for a variety of masses, namely 
$m_c=1.15-1.55~\GeV$ in 
steps of $0.05~\GeV$ and $m_b=4.25-5.25~\GeV$ in 
steps of $0.25~\GeV$. We also make available the standard MMHT2014 PDFs, and the 
sets with varied masses in the 3 and 4 flavour number schemes\footnote{}.

\section{Dependence on the heavy-quark masses}

\subsection{Choice of range of heavy-quark masses}

In the study of heavy-quark masses that accompanied the MSTW2008 PDFs 
\cite{MSTWhf} we varied the charm and beauty quark masses, defined in the 
pole mass scheme, from $m_c=1.05~\GeV$ to $m_c=1.75~\GeV$ and $m_b=4~\GeV$
to $m_b=5.5~\GeV$. This was a very generous range of masses, and it was not 
clear that there was a demand for PDFs at the extreme limits. Hence, this time 
we are a little more restrictive, and study the effects of varying $m_c$ from 
$1.15~\GeV$ to $1.55~\GeV$, in steps of $0.05~\GeV$, and of varying $m_b$ from $4.25~\GeV$
to $5.25~\GeV$ in steps of $0.25~\GeV$. Part of the reason for this is that the values are 
constrained by the comparison to data, though for both charm and beauty the
preferred values are at the lower end of the range, as we will show. However, 
there is also the constraint from other determinations of the quark masses. 
These are generally quoted in the $\MS$ scheme, and in \cite{PDG2014} 
are given as $m_c(m_c)=(1.275 \pm 0.025)~\GeV$ and 
$m_b(m_b)=(4.18\pm 0.03)~\GeV$. The transformation to the pole mass definition 
is not well-defined  due to the diverging series, i.e. there is a renormalon 
ambiguity of $\sim 0.1-0.2~\GeV$. The series is less convergent for the 
charm quark, due to the lower scale in the coupling, but the renormalon 
ambiguity cancels in difference between the charm and beauty masses. Indeed,
we obtain $m_b^{\rm pole}-m_c^{\rm pole}=3.4~\GeV$ with a very small uncertainty
\cite{Bauer:2004ve,Hoang:2005zw}. Using the perturbative expression for the 
conversion of the beauty mass, and the relationship between the beauty and 
charm mass, as shown in \cite{MSTWhf}, we obtain 
\be
m_c^{\rm pole}=1.5 \pm 0.2 ~{\GeV}  ~~~~ {\rm and} ~~~~ m_b^{\rm pole}=4.9\pm 0.2~\GeV. 
\ee 
This disfavours $m_c\leq 1.2-1.3~\GeV$ and $m_b\leq 4.6-4.7~\GeV$. As the fit
quality prefers values in this region, or lower, we allow some values a little
lower than this. In the upper direction the fit quality clearly deteriorates, 
so our upper values are not far beyond the central values quoted above. 
There is some indication from PDF fits for
a slightly lower $m^{\rm pole}$ than that suggested by the use of the perturbative series out 
to the order at which it starts to show lack of convergence. We now consider 
the variation with $m_c$ and $m_b$ in more detail. 

\begin{figure} [h]
\begin{center}
\vspace*{-0.0cm}
\includegraphics[height=6cm]{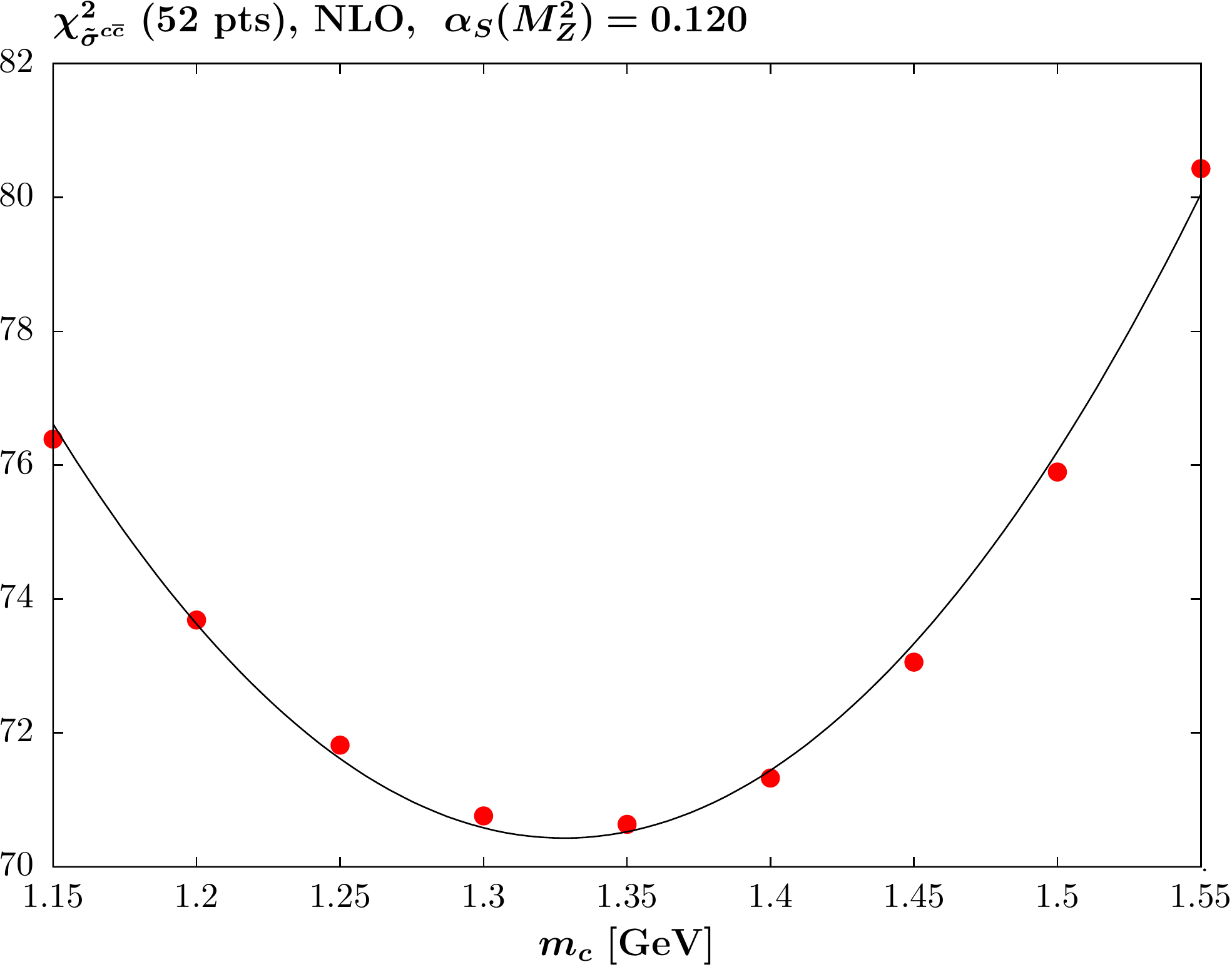}
\includegraphics[height=6cm]{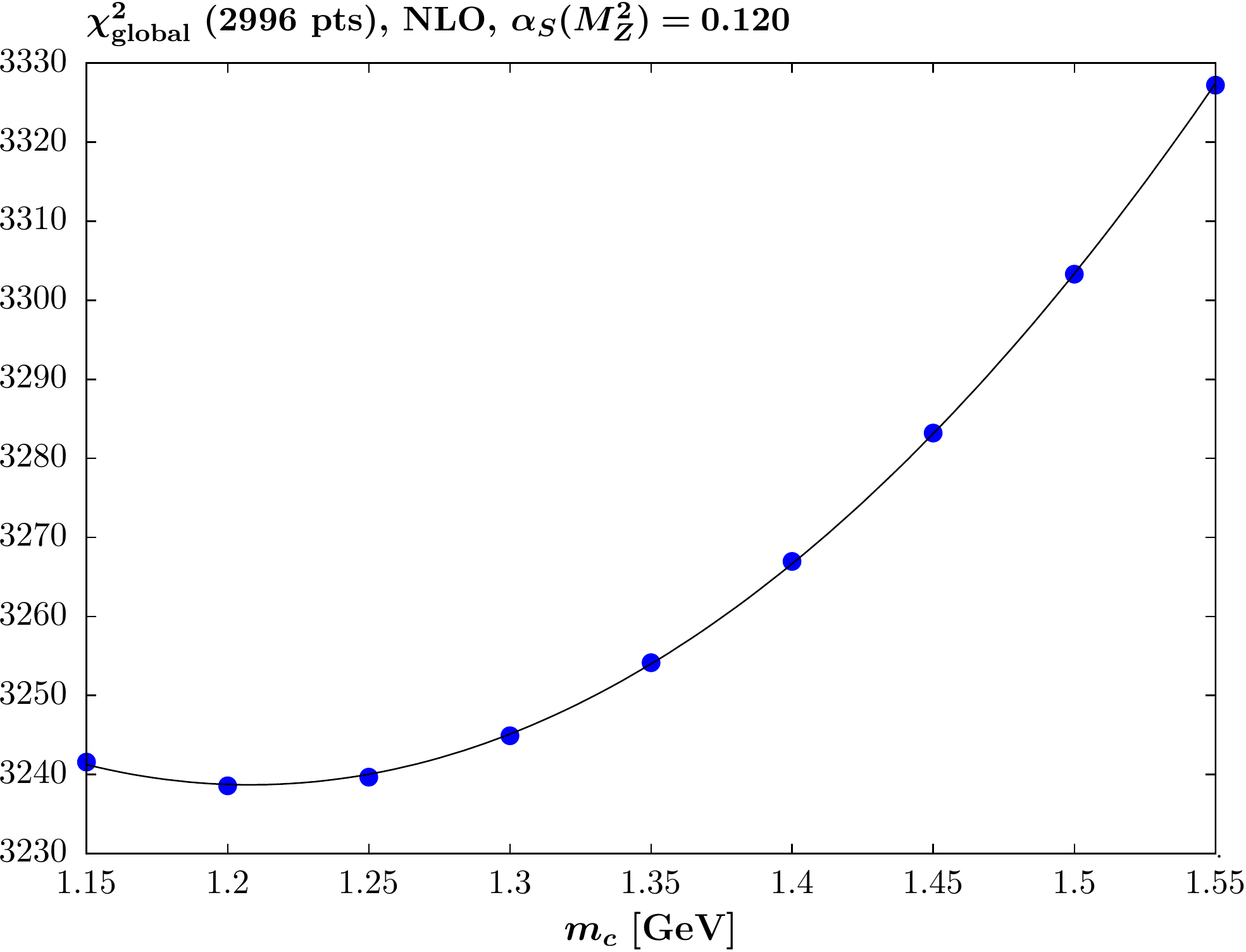}
\caption{\sf The quality of the fit versus the quark mass $m_c$ at NLO with $\alpha_S(M_Z^2)=0.120$ for (left) the reduced cross section for charm production
$\tilde{\sigma}^{\cc}$ for the combined H1 and ZEUS data and (right) the full global fit.}
\label{fig:mcas120}
\end{center}
\end{figure}

\subsection{Dependence on $m_c$}

We repeat the global analysis in \cite{MMHT} for values of $m_c=1.15-1.55~\GeV$
in steps of $0.05~\GeV$. As in \cite{MMHT} we use the ``optimal'' version    
\cite{Thorne} of the TR' general mass variable flavour number scheme 
GM-VFNS \cite{TR1}. This 
is smoother near the transition point, which we define to be at 
$Q^2=\mu^2 =m_c^2$, than the original version, so
has a slight tendency to prefer lower masses - the older version growing a 
little more quickly at low scales, which could be countered by increasing the mass. We also assume all heavy flavour is generated by evolution from the gluon
and light quarks, i.e. there is no intrinsic heavy flavour.  
We perform the analysis with $\alpha_S(M_Z^2)$ left as a free parameter in 
the fit at both NLO and NNLO, but also use our fixed default values of the 
coupling of $\alpha_S(M_Z^2)=0.118$ and 0.120 at NLO and $\alpha_S(M_Z^2)=0.118$
at NNLO. Unlike the MSTW2008 study \cite{MSTWhf} we will concentrate on the 
results and PDFs with fixed coupling, as 
the standard MMHT PDFs were made available at these values.      

\begin{figure} [h]
\begin{center}
\vspace*{-0.0cm}
\includegraphics[height=6cm]{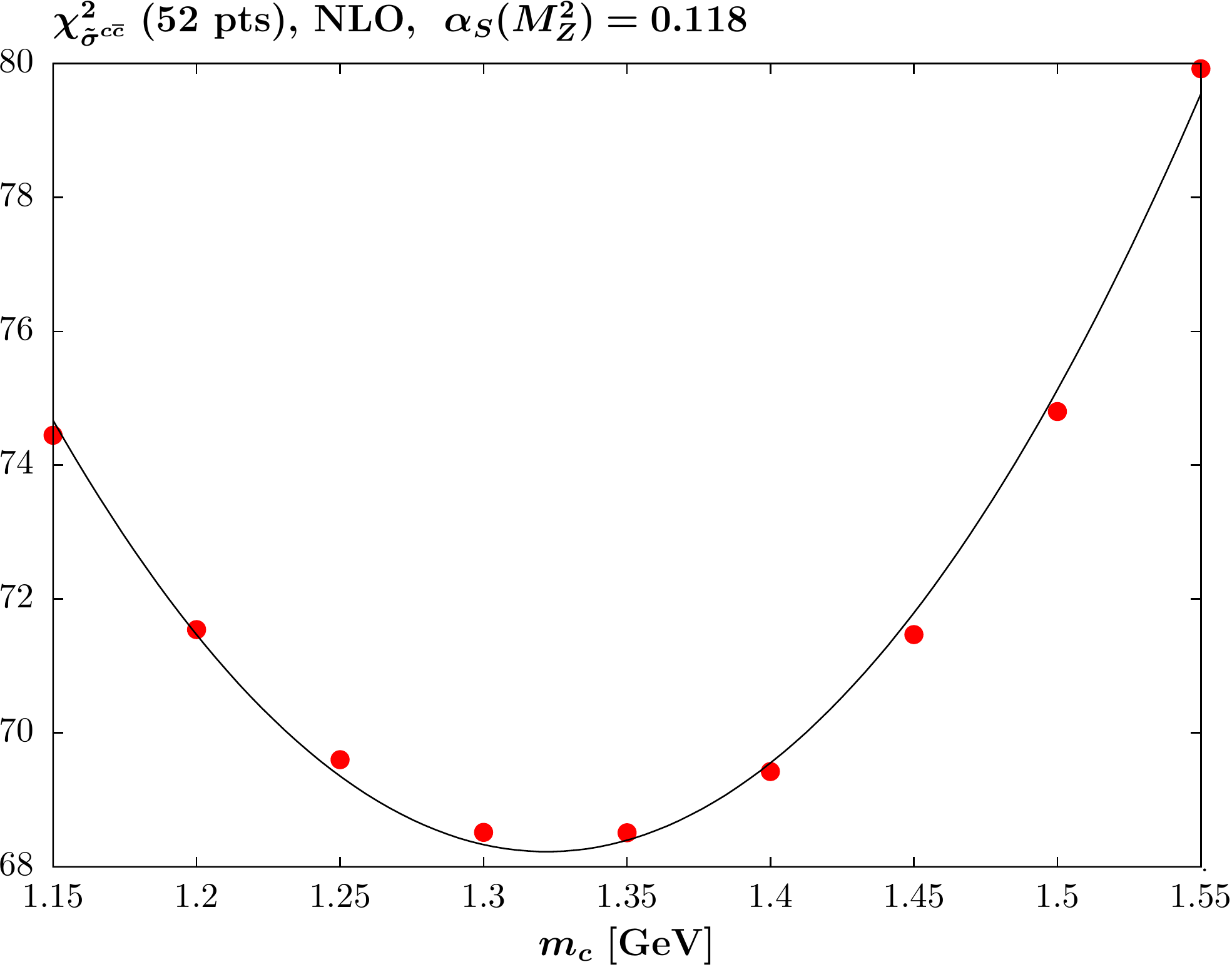}
\includegraphics[height=6cm]{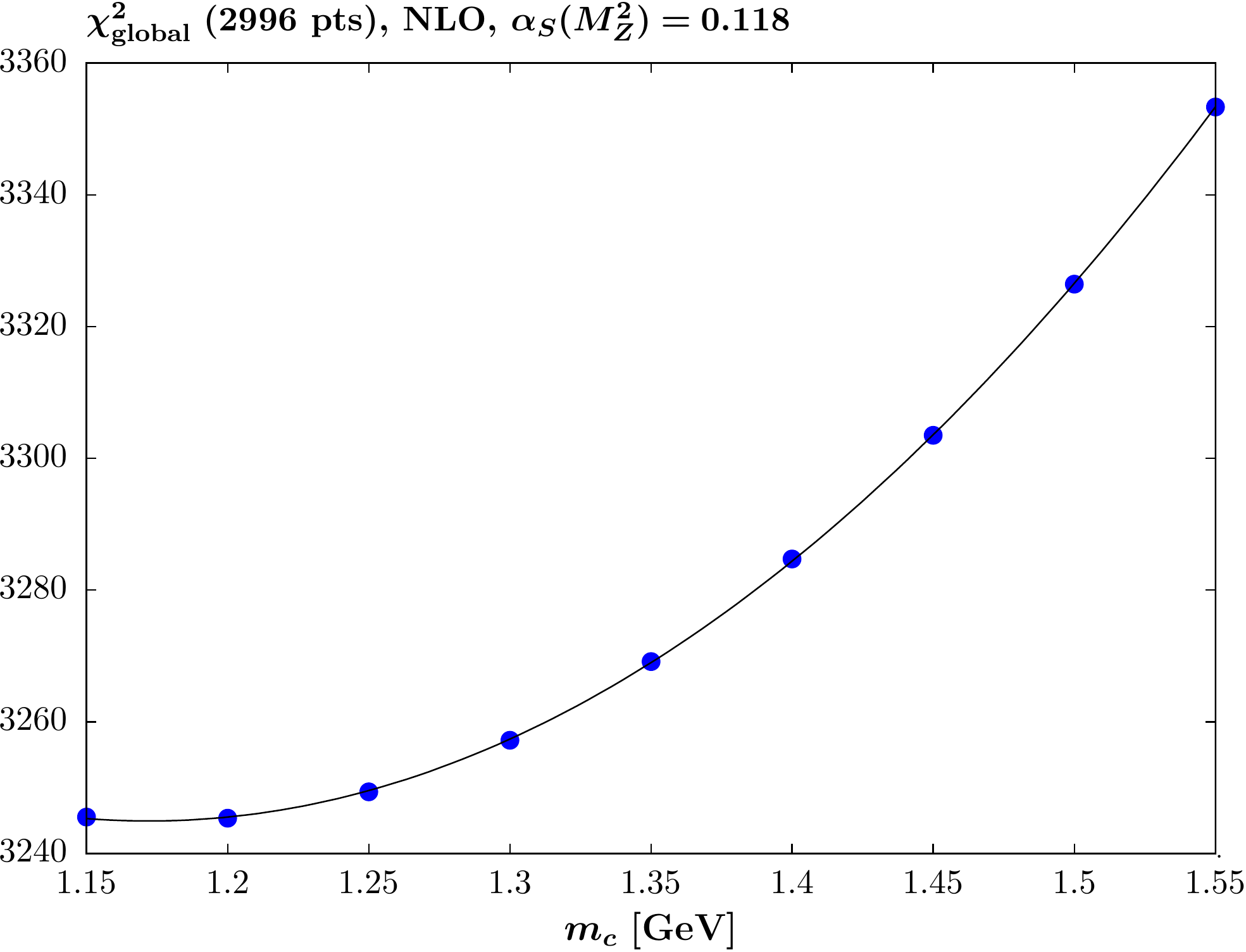}
\caption{\sf The quality of the fit versus the quark mass $m_c$ at NLO with $\alpha_S(M_Z^2)=0.118$ for (left) the reduced cross section for charm production
$\tilde{\sigma}^{\cc}$ for the combined H1 and ZEUS data and (right) the full global fit.}
\label{fig:mcas118}
\end{center}
\end{figure}

We present results in terms of the $\chi^2$ for the total set of data
in the global fit and for just the data on the reduced cross section, $\tilde{\sigma}^{\cc}$, for open 
charm production at HERA \cite{H1+ZEUScharm}. This is shown at NLO with 
$\alpha_S(M_Z^2)=0.120$ in Fig.~\ref{fig:mcas120}. The variation in the  
quality of the fit to the HERA {\it combined} charm cross section data is relatively
slight, less than the variation in the fit to the {\it separate} H1 and ZEUS data used in 
\cite{MSTWhf}. This is presumably due to the use of the full information now
available on correlated systematics, which allows movement of the data relative
to the theory with only a moderate penalty in $\chi^2$. The lower variation is 
also likely due in part to the improved flavour scheme. Despite the fairly 
small variation in $\chi^2$ the charm data clearly prefer a value close to 
$m_c=1.35~\GeV$, near our default value of $m_c=1.4~\GeV$. However, there is 
more variation in the fit quality to the global data set, with a clear preference
for values near to $m_c=1.2~\GeV$. The deterioration is clearly such as to 
make values of $m_c>1.5~\GeV$ strongly disfavoured. The main constraint comes 
from the inclusive HERA cross section data, but there is also some preference 
for a low value of the mass from NMC structure function data, where the 
data for $x\sim 0.01$ and $Q^2\sim 4~\GeV^2$ is sensitive to the turn-on of 
the charm contribution to the structure function. Overall, there is 
some element of tension between the preferred value for the global fit and 
the fit to charm data. We do not attempt to make a rigorous determination of 
the best value of the mass or its uncertainty, as provided in 
\cite{Alekhin:2012vu} for example, as we believe there are more precise and 
better controlled methods for this. However, a rough indication of the 
uncertainty could be obtained from the $\chi^2$ profiles by treating $m_c$ 
in the same manner as the standard PDF eigenvectors and applying the dynamic 
tolerance procedure. In this case the appropriate tolerance, 
obtained by assuming the charm cross section data is the dominant constraint, 
would be of the order $T=\sqrt{\Delta \chi^2}\approx 2.5$.\footnote{As 
discussed in \cite{MSTW}, for a $68\%$ confidence level uncertainty we 
insist the
fit quality to a given data set deteriorates by no more than the width 
of the $\chi^2$ distribution for $N$ points, roughly $\sqrt{N/2}$ multiplied by
the $\chi^2$ per point for the best fit. For the charm cross section data
this is $\approx \sqrt{52/2}*1.3$.}

\begin{table}
\begin{center}
\renewcommand\arraystretch{1.25}

\begin{tabular}{|l|l|l|l|}
\hline
   $m_c$ (GeV)       &  $\chi^2_{\rm global}$  &   $\chi^2_{\tilde{\sigma}^{\cc}}$ &  
$\alpha_s(M_Z^2)$        \\
          &  2996 pts &  52 pts &  \\
\hline
1.15   &  3239  & 75   & 0.1190     \\
1.2   &  3237  & 73   & 0.1192     \\
1.25   &  3239  & 71   & 0.1194     \\
1.3  & 3245  &  70  &  0.1195   \\
1.35  & 3254  &  70  &  0.1196   \\
1.4  &  3268   &  71  & 0.1198     \\
1.45  &  3283   &  73  & 0.1200     \\
1.5  & 3303  &  76  &  0.1201   \\
1.55  & 3327  &  81  &  0.1202   \\
\hline
    \end{tabular}
\end{center}

\caption{\sf The quality of the fit versus the quark mass $m_c$ at NLO with $\alpha_S(M_Z^2)$ 
left as a free parameter.}
\label{tab:mcasfree}   
\end{table}

\begin{figure} [h]
\begin{center}
\vspace*{-0.0cm}
\includegraphics[height=6cm]{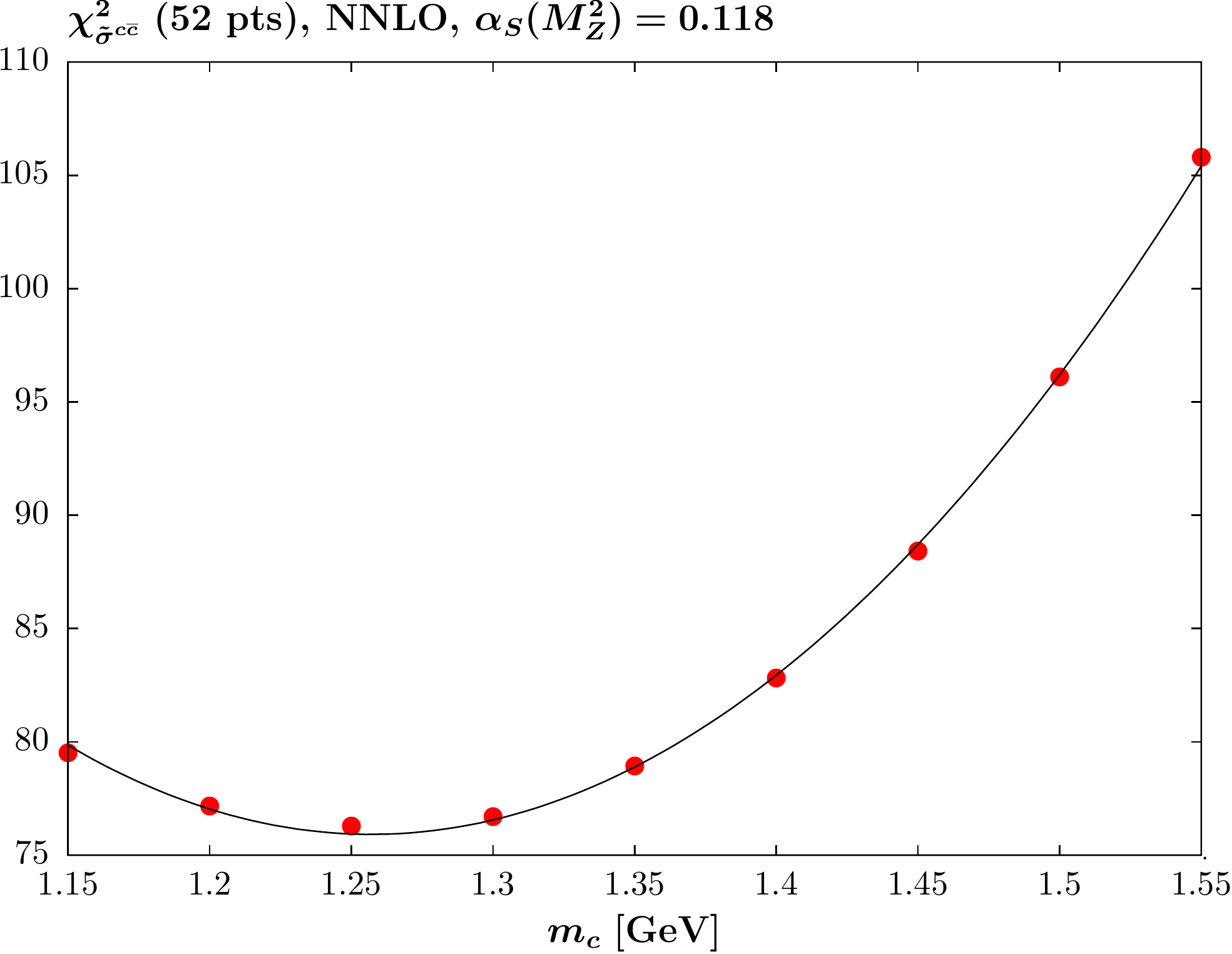}
\includegraphics[height=6cm]{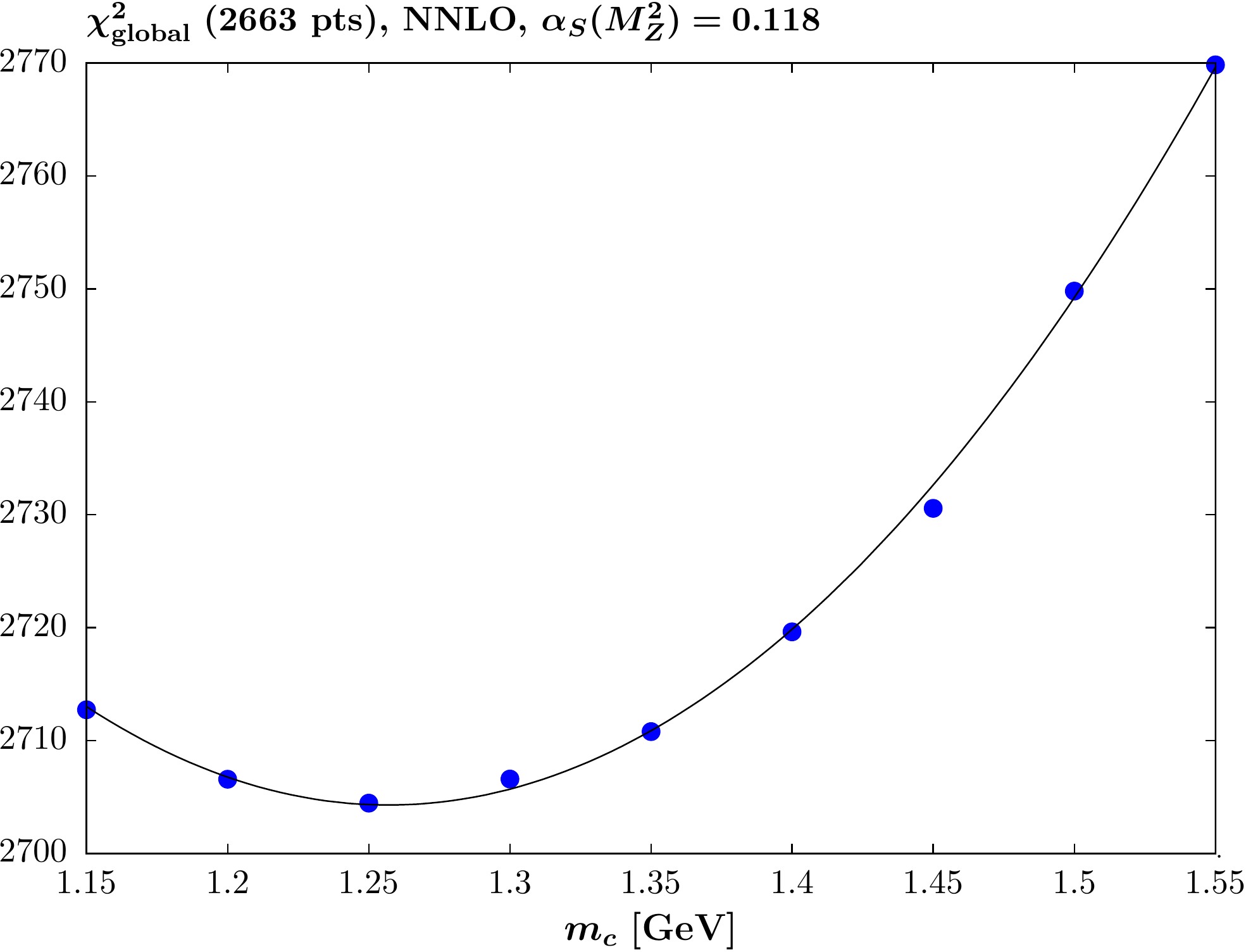}
\caption{\sf The quality of the fit versus the quark mass $m_c$ at NNLO with $\alpha_S(M_Z^2)=0.118$ for (left) the reduced cross section for charm production
$\tilde{\sigma}^{\cc}$ for the combined H1 and ZEUS data and (right) the full global fit.
}
\label{fig:mcNNLOas118}
\end{center}
\end{figure}

\begin{table}
\begin{center}
\renewcommand\arraystretch{1.25}
\begin{tabular}{|l|l|l|l|}
\hline
   $m_c$ (GeV)       &  $\chi^2_{\rm global}$  &   $\chi^2_{\tilde{\sigma}^{\cc}}$ &  
$\alpha_s(M_Z^2)$        \\
          &  2663 pts &  52 pts &  \\
\hline
1.15   &  2703  & 78   & 0.1164     \\
1.2   &  2699  & 76   & 0.1166     \\
1.25   &  2698  & 75   & 0.1167     \\
1.3  & 2701  &  76  &  0.1169   \\
1.35  & 2707  &  78  &  0.1171   \\
1.4  &  2717   &  82  & 0.1172     \\
1.45  &  2729   &  88  & 0.1173     \\
1.5  & 2749  &  96  &  0.1173   \\
1.55  & 2769  & 105  &  0.1175   \\
\hline
    \end{tabular}
\end{center}

\caption{\sf The quality of the fit versus the quark mass $m_c$ at NNLO with $\alpha_S(M_Z^2)$
left free.}
\label{tab:mcNNLOasfree}   
\end{table}

The analogous results for $\alpha_S(M_Z^2)=0.118$ and $\alpha_S(M_Z^2)$ left 
free  are shown in Fig.~\ref{fig:mcas118} and Table \ref{tab:mcasfree}
respectively, where in the latter case the corresponding $\alpha_s(M_Z^2)$ values are shown as well. For $\alpha_S(M_Z^2)=0.118$ the picture is much the same as 
for $\alpha_S(M_Z^2)=0.120$ except that the fit to charm data is marginally 
better, while the global fit is a little worse, though more-so for higher 
masses. The results with free  $\alpha_S(M_Z^2)$ are consistent with this, 
with the preferred value of $\alpha_S(M_Z^2)$ falling slightly with 
lower values of $m_c$. However, the values of $m_c$ preferred by charm data and
the full data sets are much the same as for fixed coupling --- the values of the 
$\chi^2$ just being a little lower in general.

The results of the same analysis at NNLO are shown for $\alpha_S(M_Z^2)=0.118$
and $\alpha_S(M_Z^2)$ left free in Fig.~\ref{fig:mcNNLOas118} and Table
\ref{tab:mcNNLOasfree}, respectively, where again  in the latter case the corresponding $\alpha_s(M_Z^2)$ values are shown. Broadly speaking, the results are similar to those at NLO, but with 
lower values of $m_c$ preferred and where the $\chi^2$ variation is greater for the 
inclusive data than for the charm cross section data. However, in this case 
there is essentially no tension at all between the inclusive and charm data, 
with both $\chi^2$ values minimising very near to $m_c=1.25~\GeV$ --- this lower 
preferred value for the charm data meaning that the fit quality at $m_c=1.55~\GeV$ has
deteriorated more than at NLO. The picture is exactly the same for fixed and 
free strong coupling, with the values of $\chi^2$ simply being a little lower 
when $\alpha_S(M_Z^2)$ is left free, since the best fit value of the coupling is 
a little below $0.118$, particularly for low $m_c$.

\subsection{Dependence on $m_b$}

\begin{figure} [h]
\begin{center}
\vspace*{-0.0cm}
\includegraphics[height=6cm]{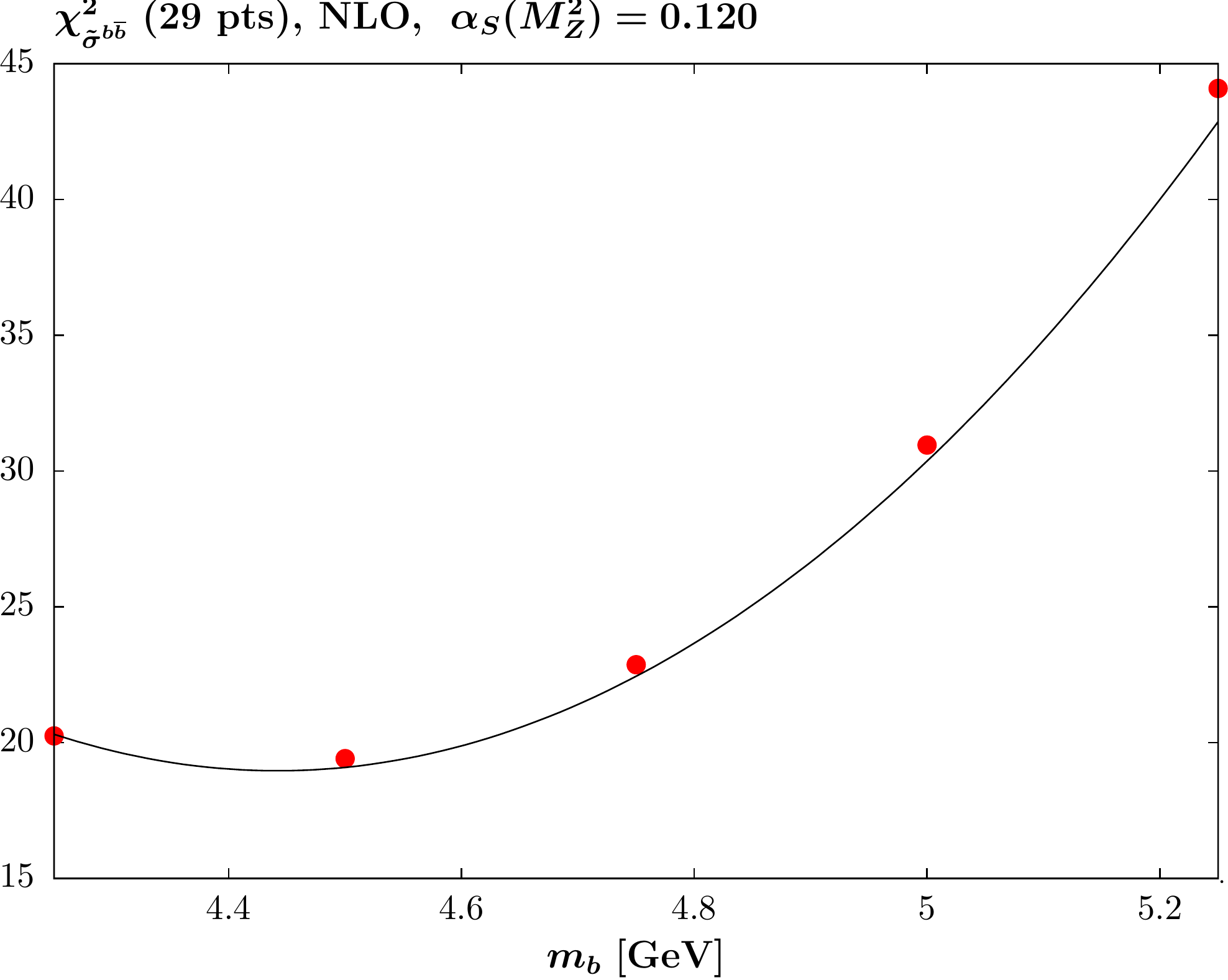}
\includegraphics[height=6cm]{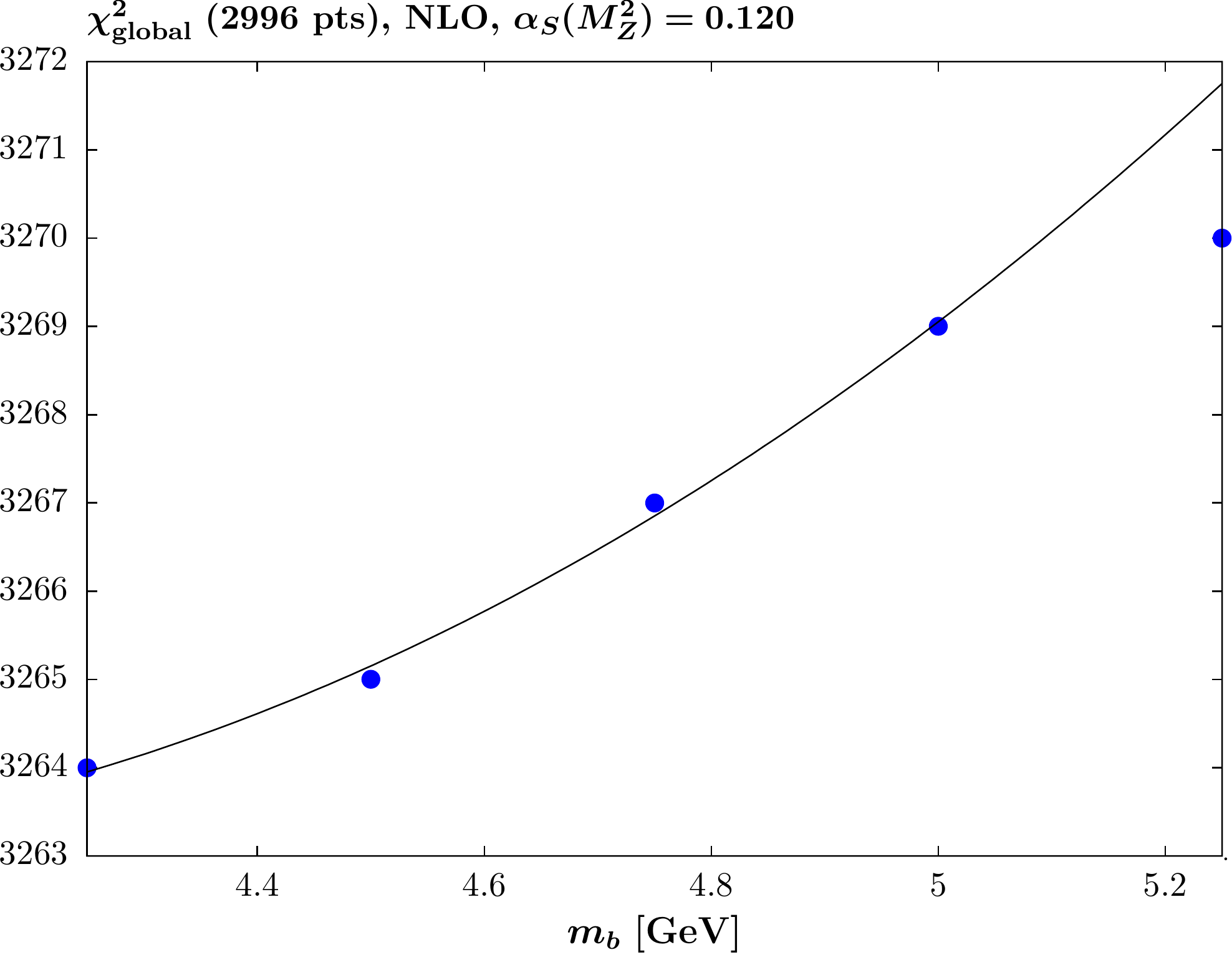}
\caption{\sf The quality of the fit versus the quark mass $m_b$ at NLO with $\alpha_S(M_Z^2)=0.120$ for (left) the reduced cross section for beauty production
$\tilde{\sigma}^{\bb}$ for the H1 and ZEUS data and (right) the global fit, not including 
the beauty data.}
\label{fig:mbas120}
\end{center}
\end{figure}

We repeat essentially the same procedure for varying values of 
$m_b$ in the range $4.25-5.25~\GeV$ in steps of $0.25~\GeV$. However, this time there were
no data on the beauty contribution to the cross section included in the standard
global fit \cite{MMHT}. In the previous heavy-quark analysis \cite{MSTWhf} we compared 
to beauty cross section data from H1~\cite{Aaron:2009af}. This placed a weak 
constraint on the value of $m_b$ but had negligible constraint on the PDFs
for fixed $m_b$. Hence, we did not include these data in the updated global 
fit \cite{MMHT}. There are now also data of comparable precision from ZEUS 
\cite{Abramowicz:2014zub}, and we will include both these data sets in 
future global fits. In this article we study the quality of the comparison 
to these data to predictions obtained using the MMHT PDFs with different values of $m_b$. The data 
themselves are not included in the fit, i.e. we use predictions from the PDFs, 
as they still provide negligible direct constraint.

\begin{figure} [h]
\begin{center}
\vspace*{-0.0cm}
\includegraphics[height=6cm]{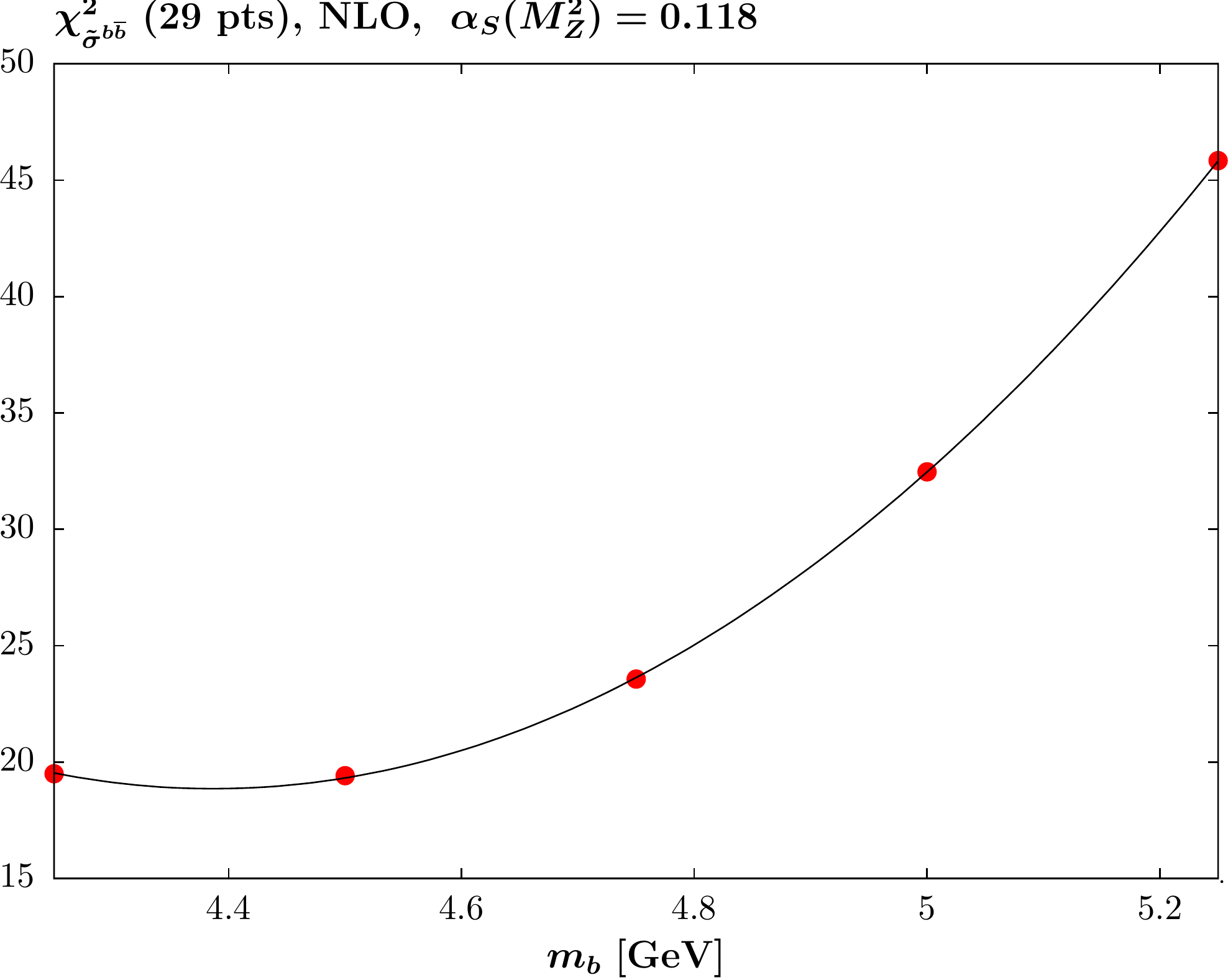}
\includegraphics[height=6cm]{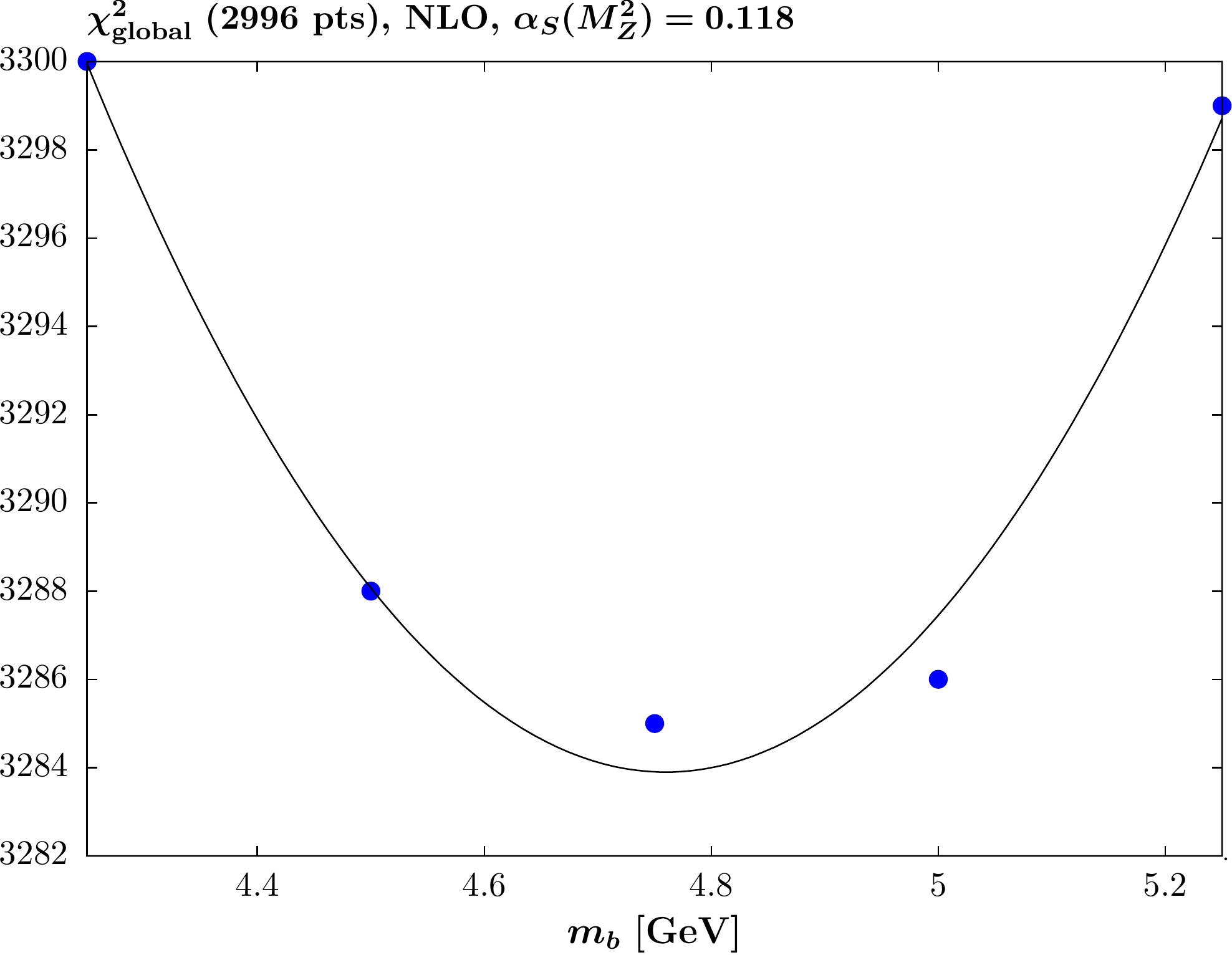}
\caption{\sf The quality of the fit versus the quark mass $m_b$ at NLO with $\alpha_S(M_Z^2)=0.118$ for (left) the reduced cross section for beauty production
$\tilde{\sigma}^{\bb}$ for the H1 and ZEUS data and (right) the global fit, not including 
the beauty data.}
\label{fig:mbas118}
\end{center}
\end{figure}

\begin{figure} [h]
\begin{center}
\vspace*{-0.0cm}
\includegraphics[height=6cm]{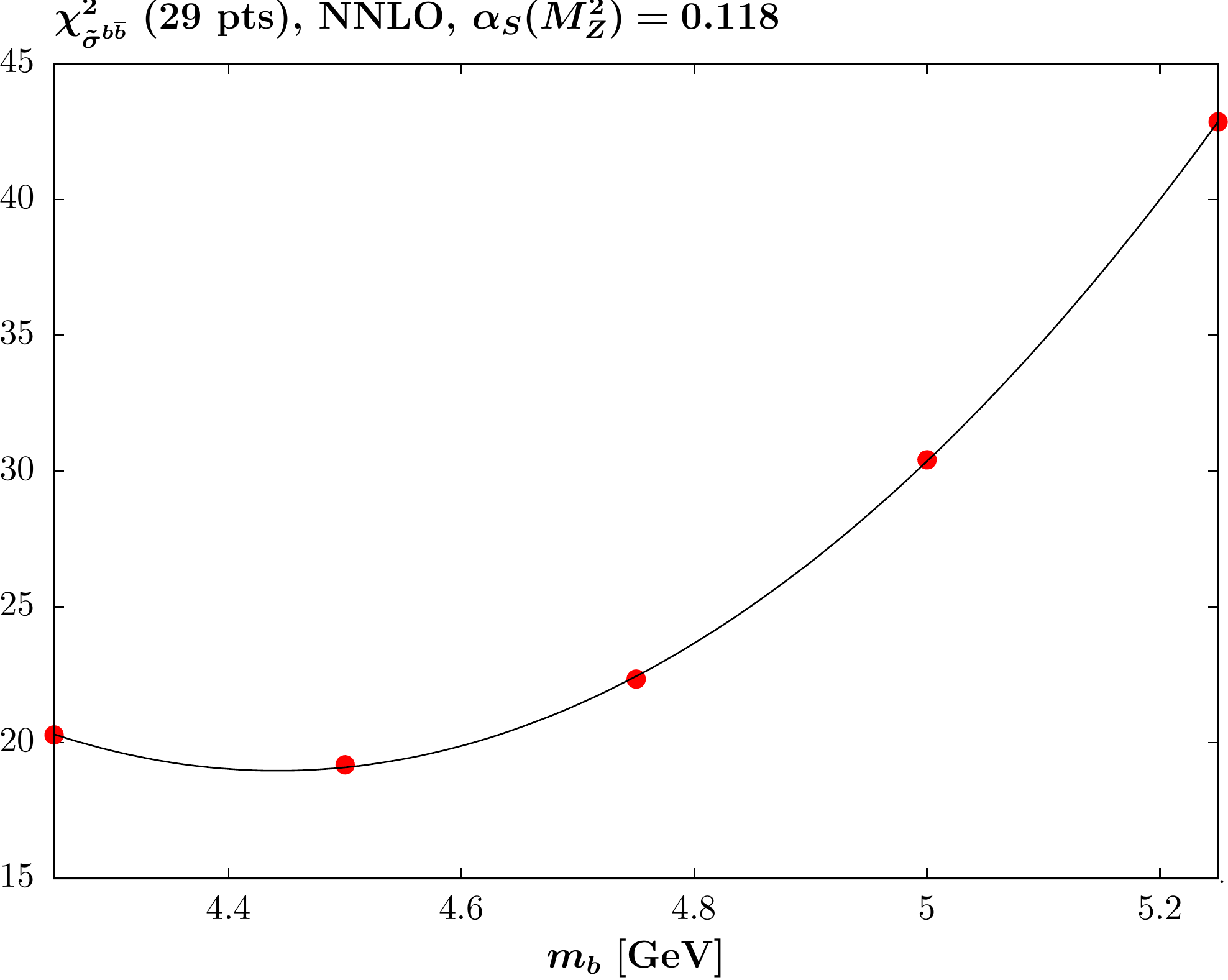}
\includegraphics[height=6cm]{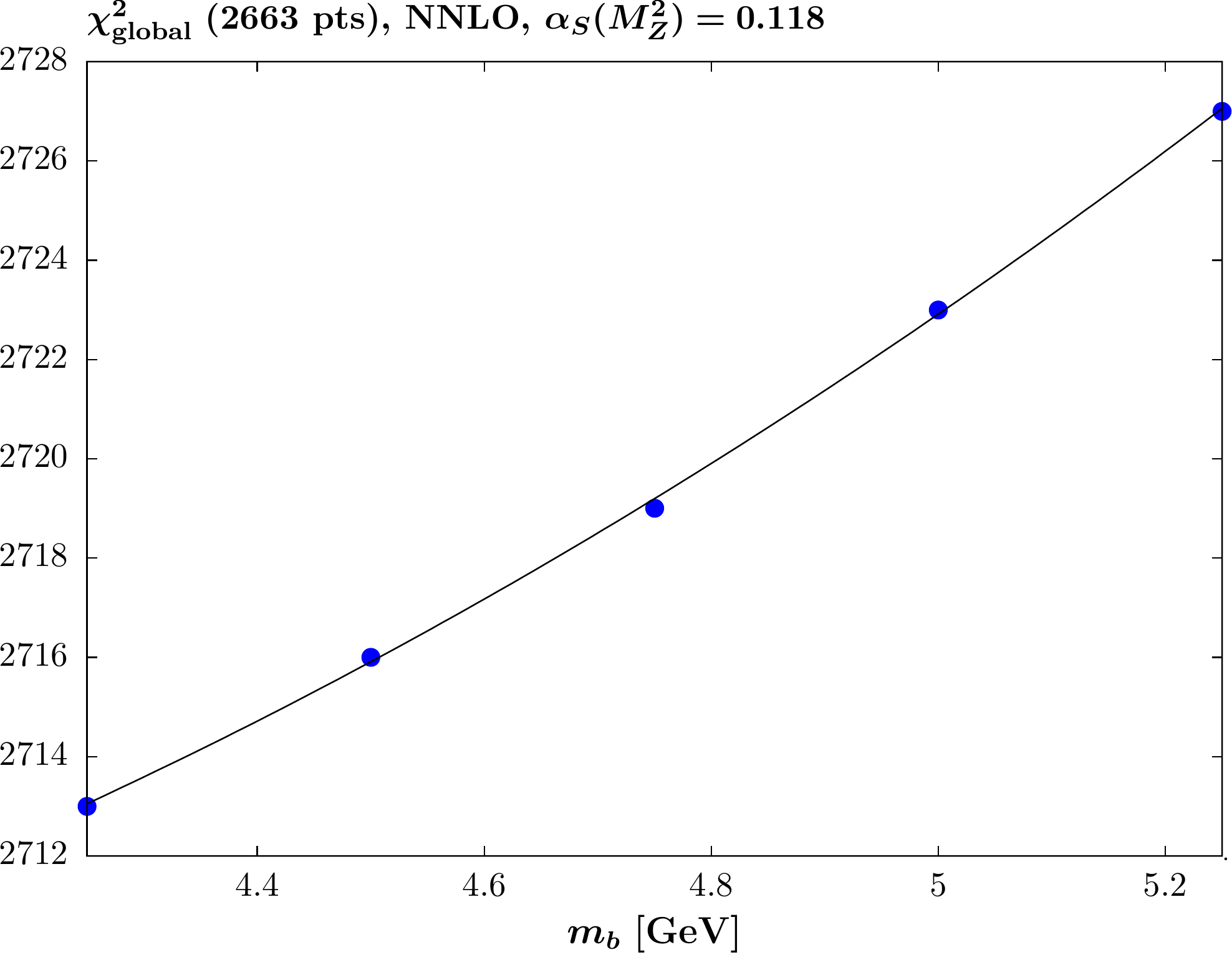}
\includegraphics[height=6cm]{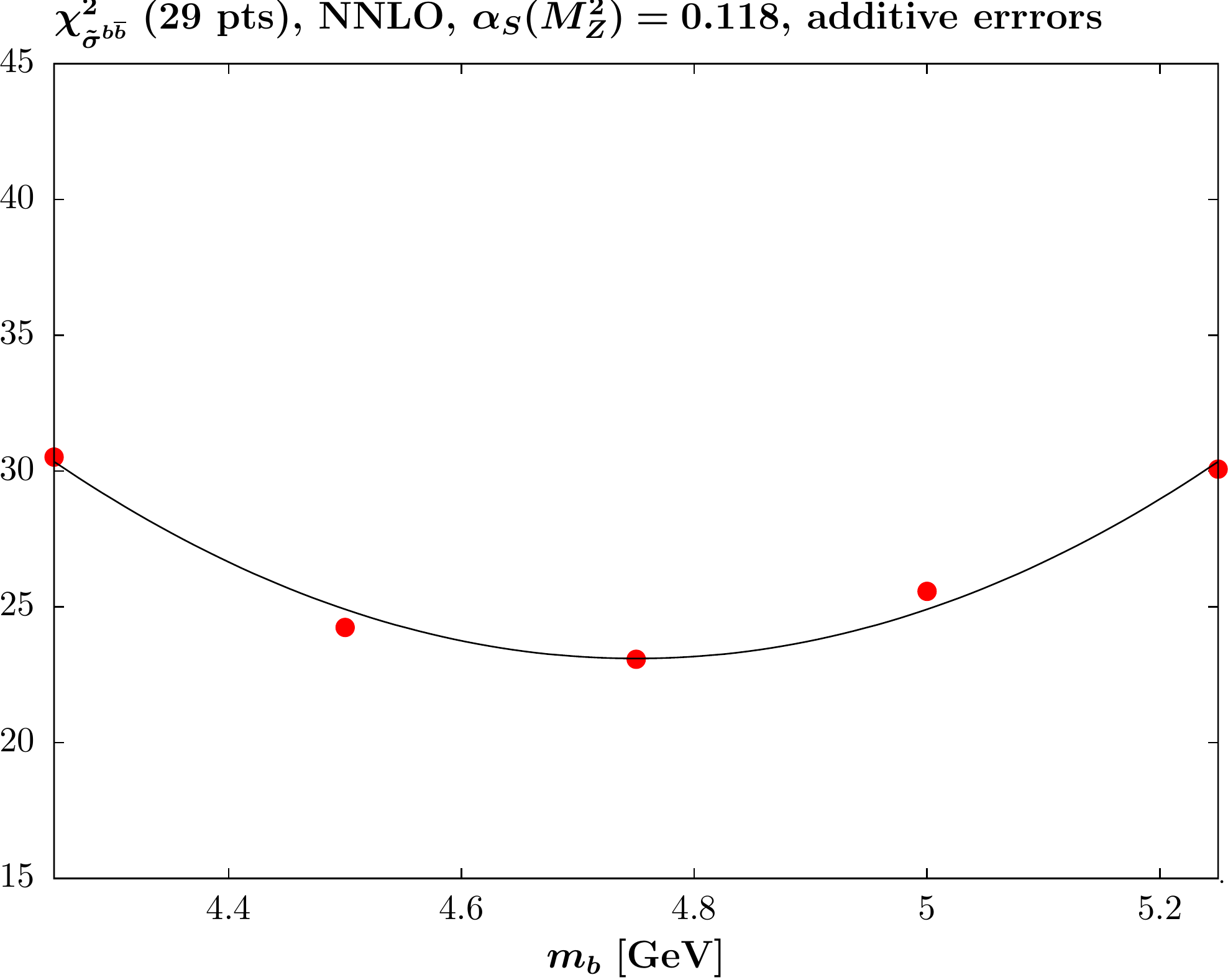}
\caption{\sf The quality of the fit versus the quark mass $m_b$ at NNLO with $\alpha_S(M_Z^2)=0.118$ for (left) the reduced cross section for beauty production
$\tilde{\sigma}^{\bb}$ for the H1 and ZEUS data and (right) the global fit, not including 
the beauty data.
Recall that in the MMHT analysis the experimental errors are treated multiplicatively. The lower plot shows the $\chi^2$ profile if the errors in the HERA beauty data were to be treated additively.}
\label{fig:mbasNNLO118}
\end{center}
\end{figure}

The results for the NLO PDFs with $\alpha_S(M_Z^2)=0.120$ and 0.118 
are shown in Figs.~\ref{fig:mbas120} and ~\ref{fig:mbas118} respectively. The picture for the data in the global fit (not including the $\tilde{\sigma}^{\bb}$ data)
is slightly 
different in the two cases: for $\alpha_S(M_Z^2)=0.120$ there is a fairly weak 
tendency to prefer lower values of $m_b$, similar to the results in 
\cite{MSTWhf}, but for $\alpha_S(M_Z^2)=0.118$ the global fit prefers a value
of between $4.5~\GeV$ and $5.0~\GeV$. For the predictions for the beauty
cross section data, however, the picture is similar in the two cases,
and low values of $m_b \sim 4.4-4.5~\GeV$ are preferred.

The results for the NNLO fit with $\alpha_S(M_Z^2)=0.118$ are shown in Fig.~\ref{fig:mbasNNLO118}. As can be seen the global fit is fairly weakly dependent 
on $m_b$, though more than for $\alpha_S(M_Z^2)=0.120$ at NLO, 
and prefers a value lower than $m_b=4.25~\GeV$. As in the NLO case the $\chi^2$ for the 
prediction for $\tilde{\sigma}^{\bb}$ is better for lower values of $m_b$. The slightly 
larger variations in the quality of the global fit with varying $m_b$ compared
to \cite{MSTWhf} is perhaps due to the greater precision of the inclusive 
HERA cross section data used in this analysis, and to the fact that 
the CMS double-differential Drell-Yan data \cite{CMS-ddDY} has some sensitivity
to the value of $m_b$ due to the induced variation in sea quark flavour 
composition for low scales. The previous analysis preferred a value of $m_b \sim
4.75~\GeV$ for the comparison to the H1 beauty data. However, the definition
of the general mass variable number scheme has improved since this previous 
analysis, being smoother near to the transition point $Q^2=m_b^2$, 
and including an improvement to the approximation for the 
${\cal O}(\alpha_S^3)$ contribution at low $Q^2$ at NNLO, so some changes are 
not surprising. Another important difference is in the treatment of 
the correlated experimental errors, which we now take as being multiplicative. The result within exactly the same framework, but with the experimental errors on the HERA beauty data instead treated as additive is also shown in Fig.~\ref{fig:mbasNNLO118} and a higher value of $m_b \sim 4.75~\GeV$ is clearly preferred. Similar results are seen in the NLO fits. 

\begin{figure} 
\begin{center}
\vspace*{-1.0cm}
\includegraphics[scale=0.25]{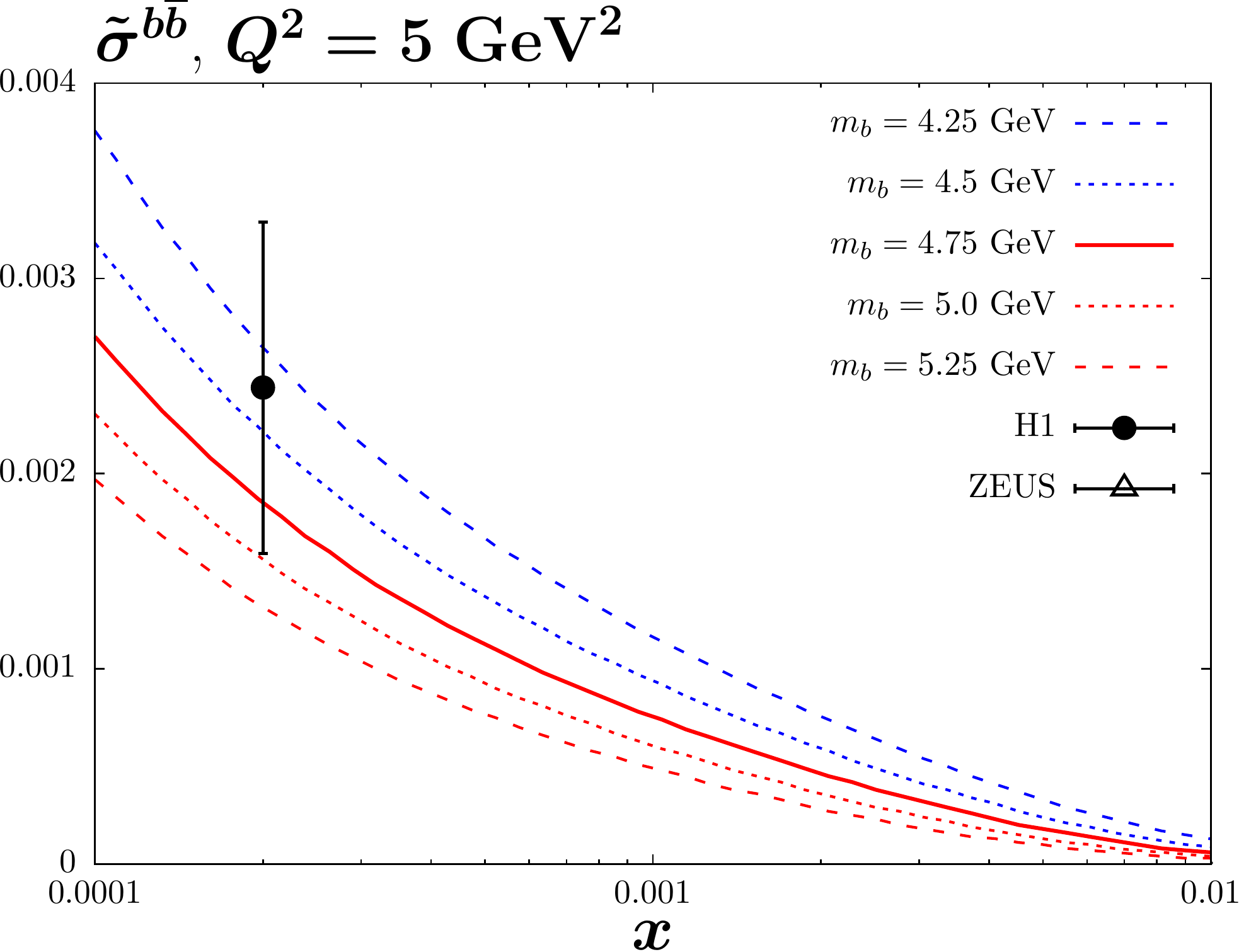}
\includegraphics[scale=0.25]{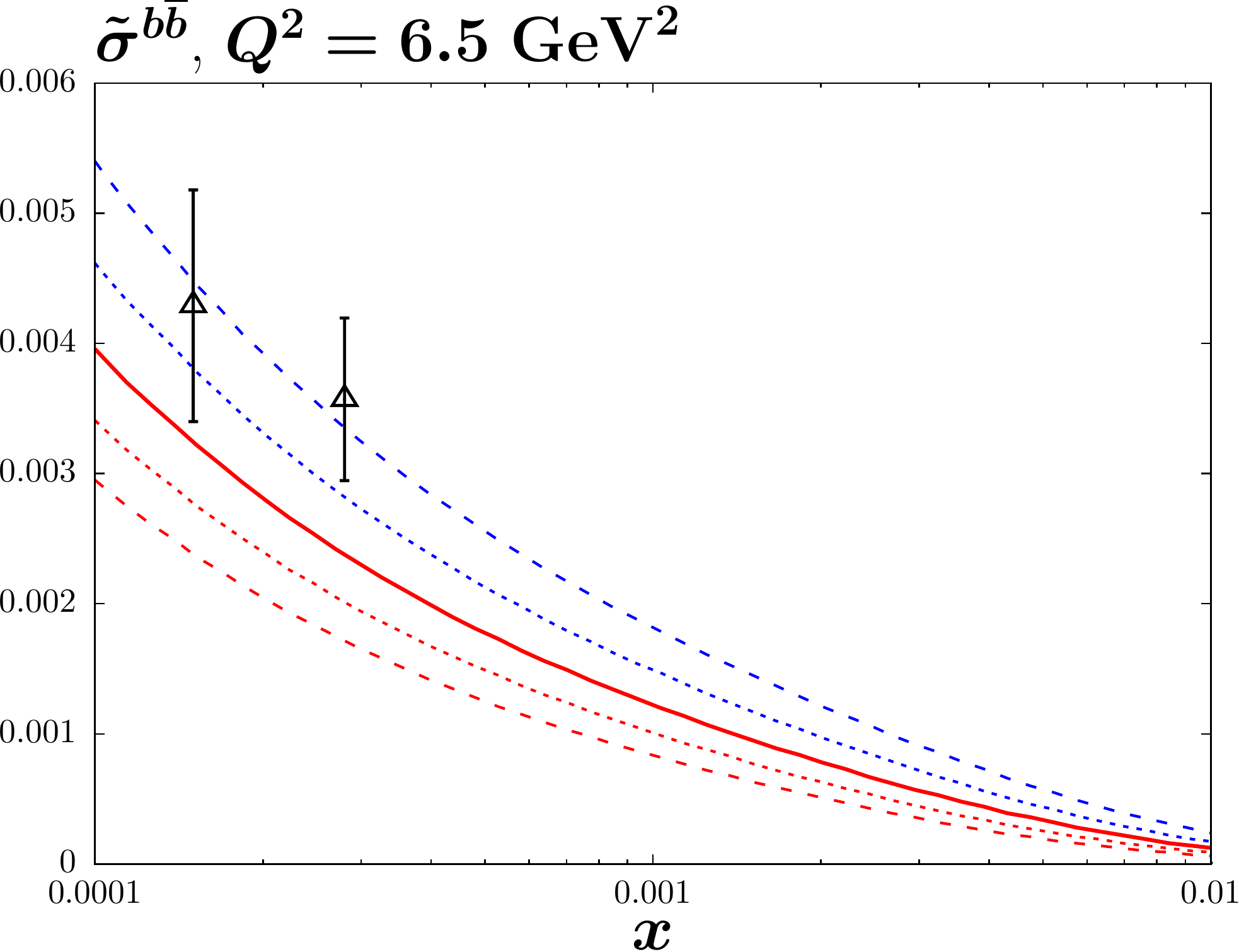}
\includegraphics[scale=0.25]{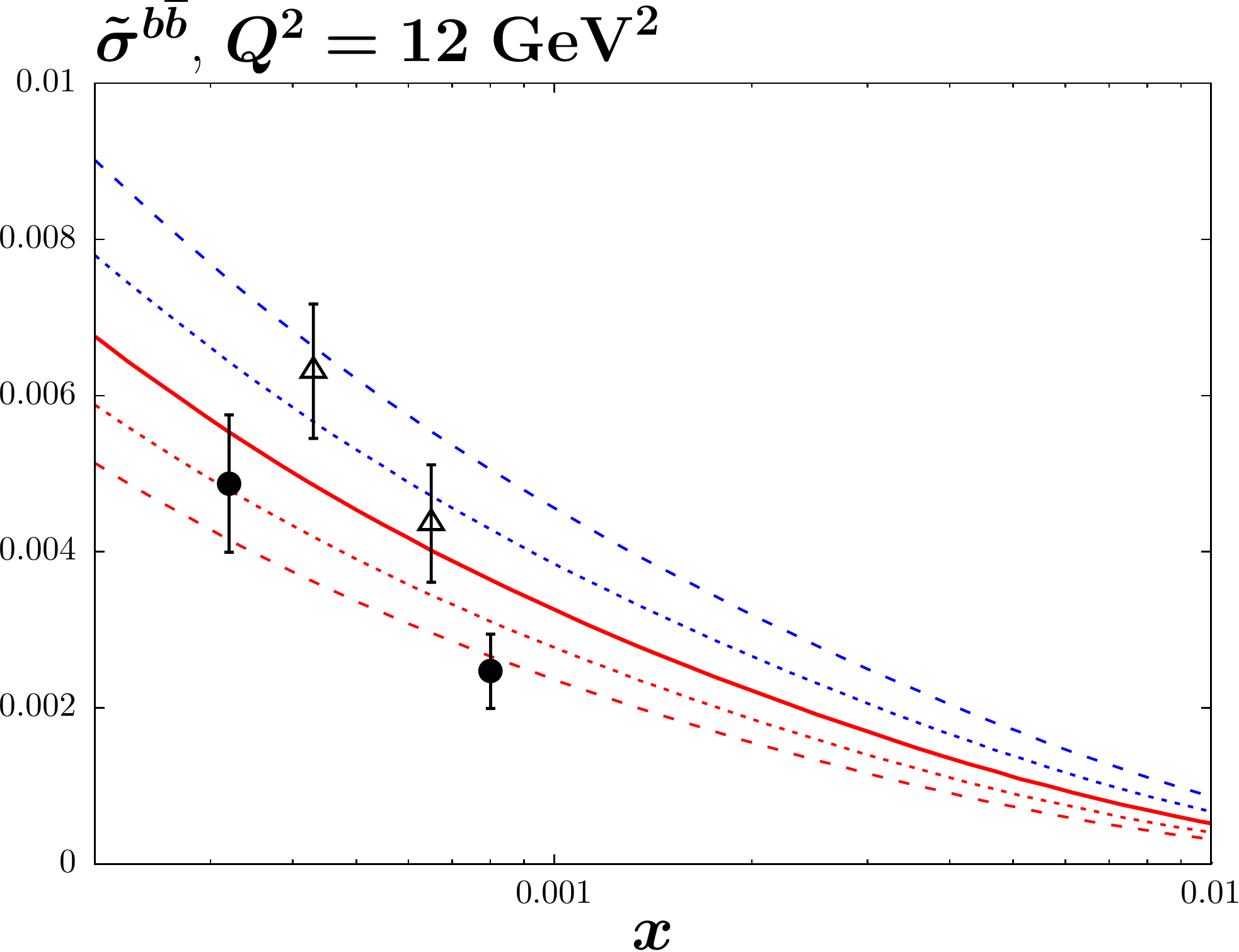}
\includegraphics[scale=0.25]{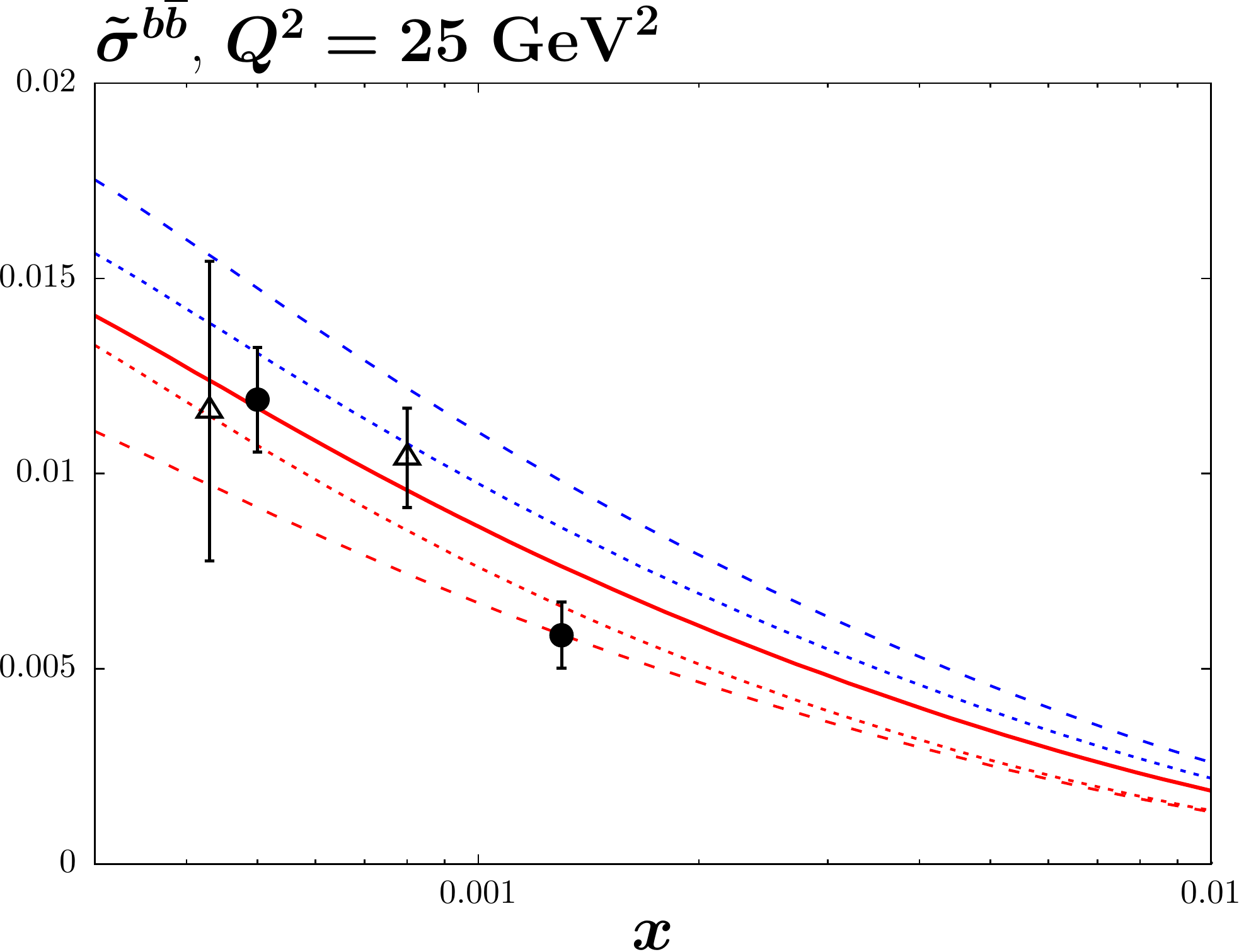}
\includegraphics[scale=0.25]{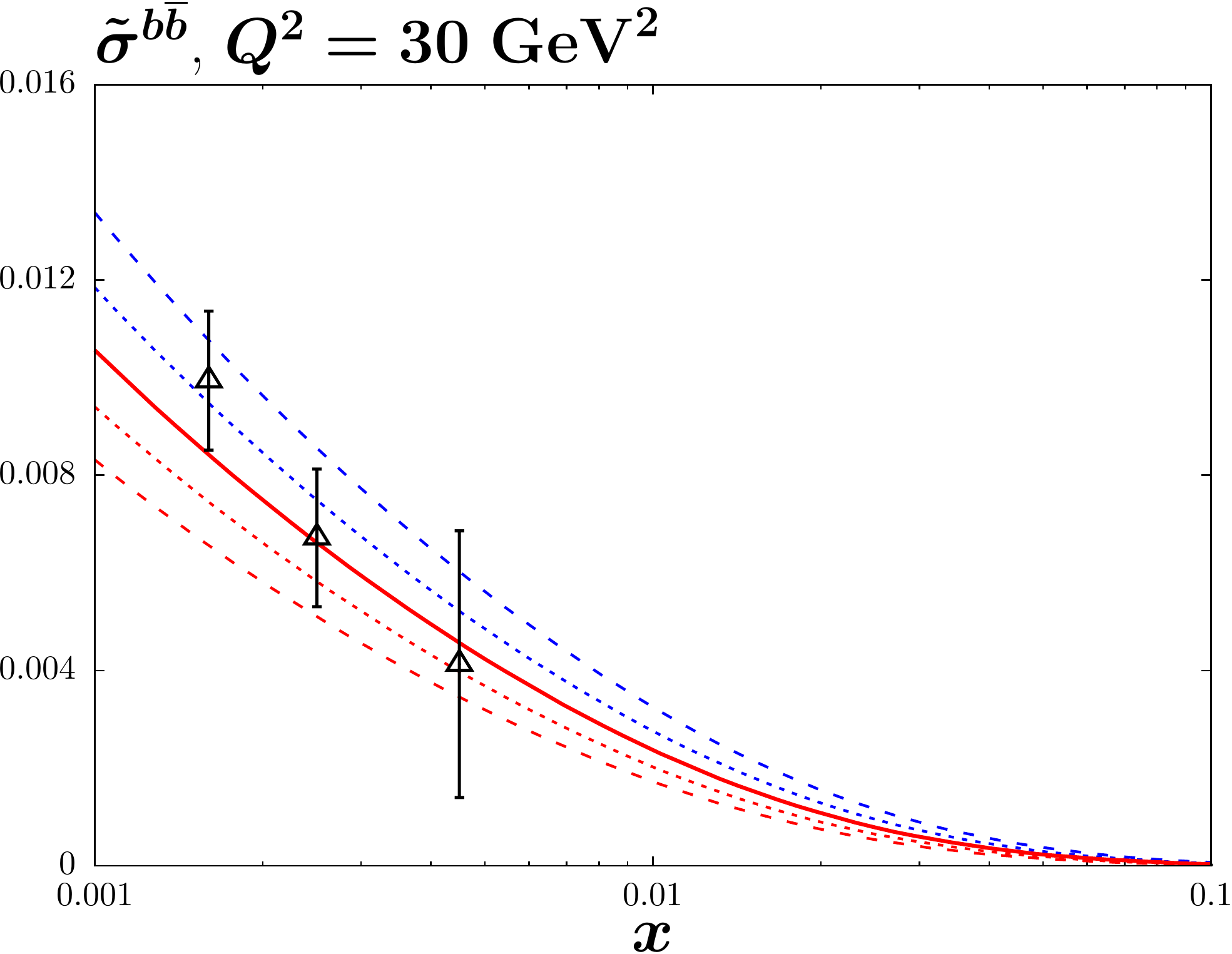}
\includegraphics[scale=0.25]{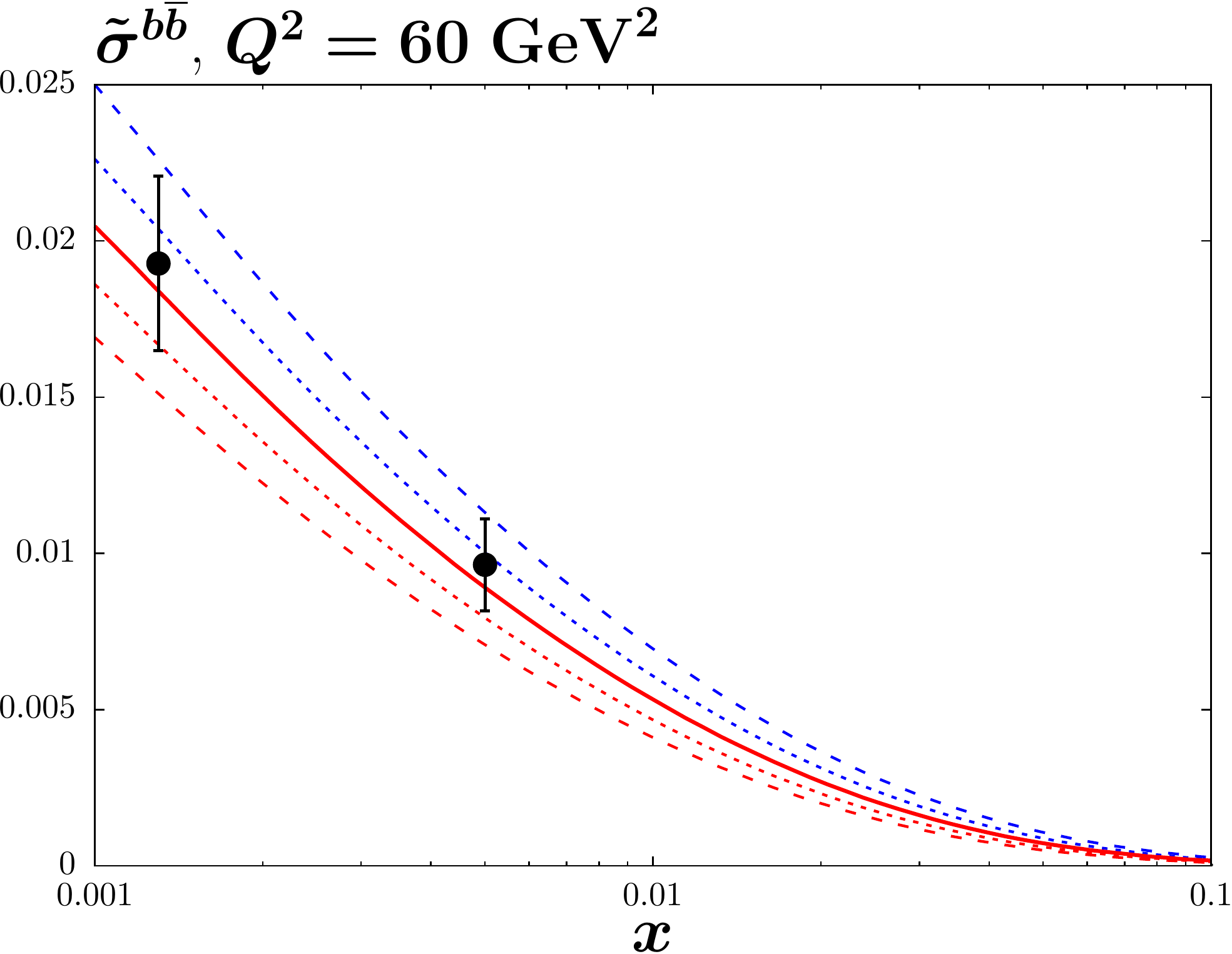}
\includegraphics[scale=0.25]{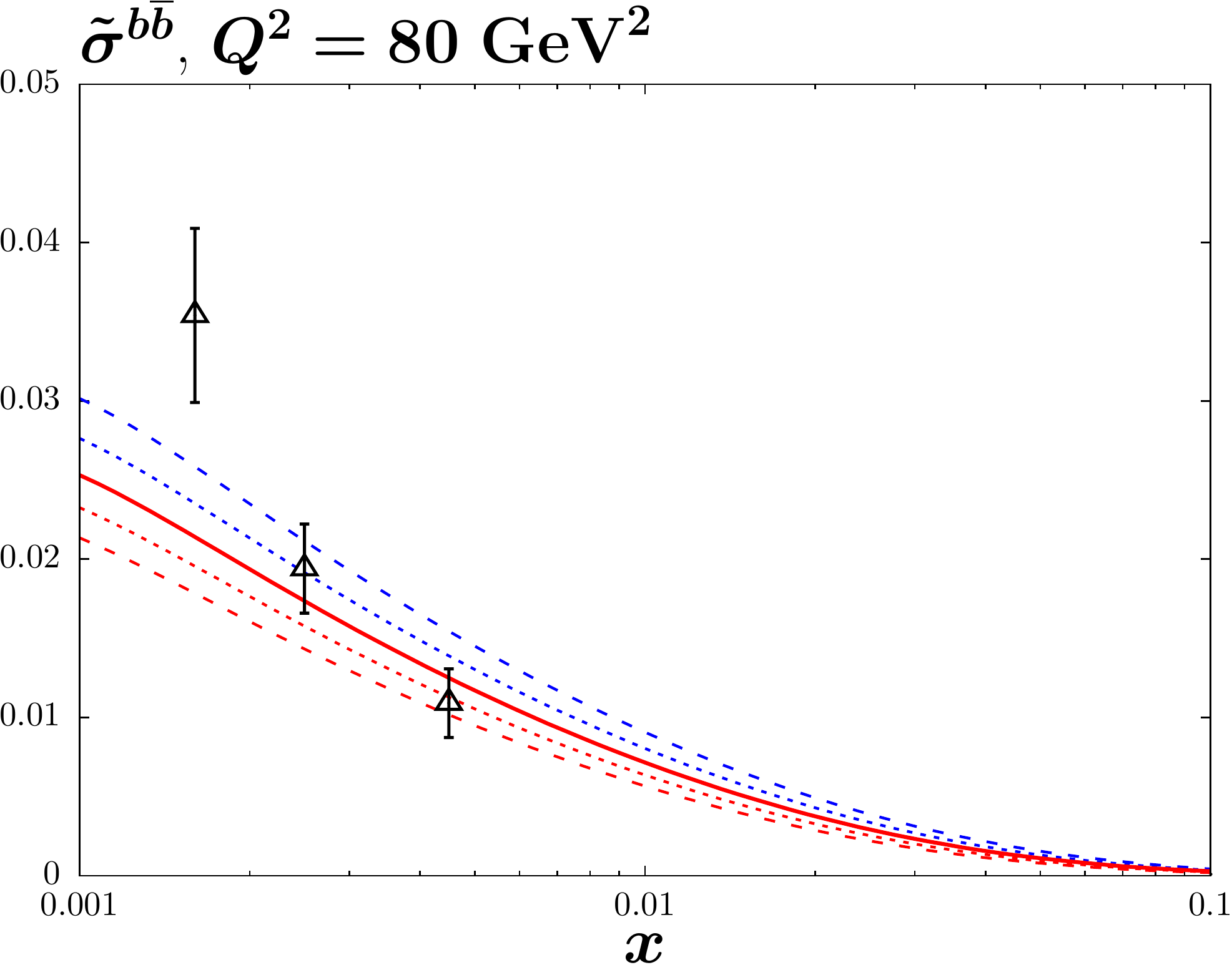}
\includegraphics[scale=0.25]{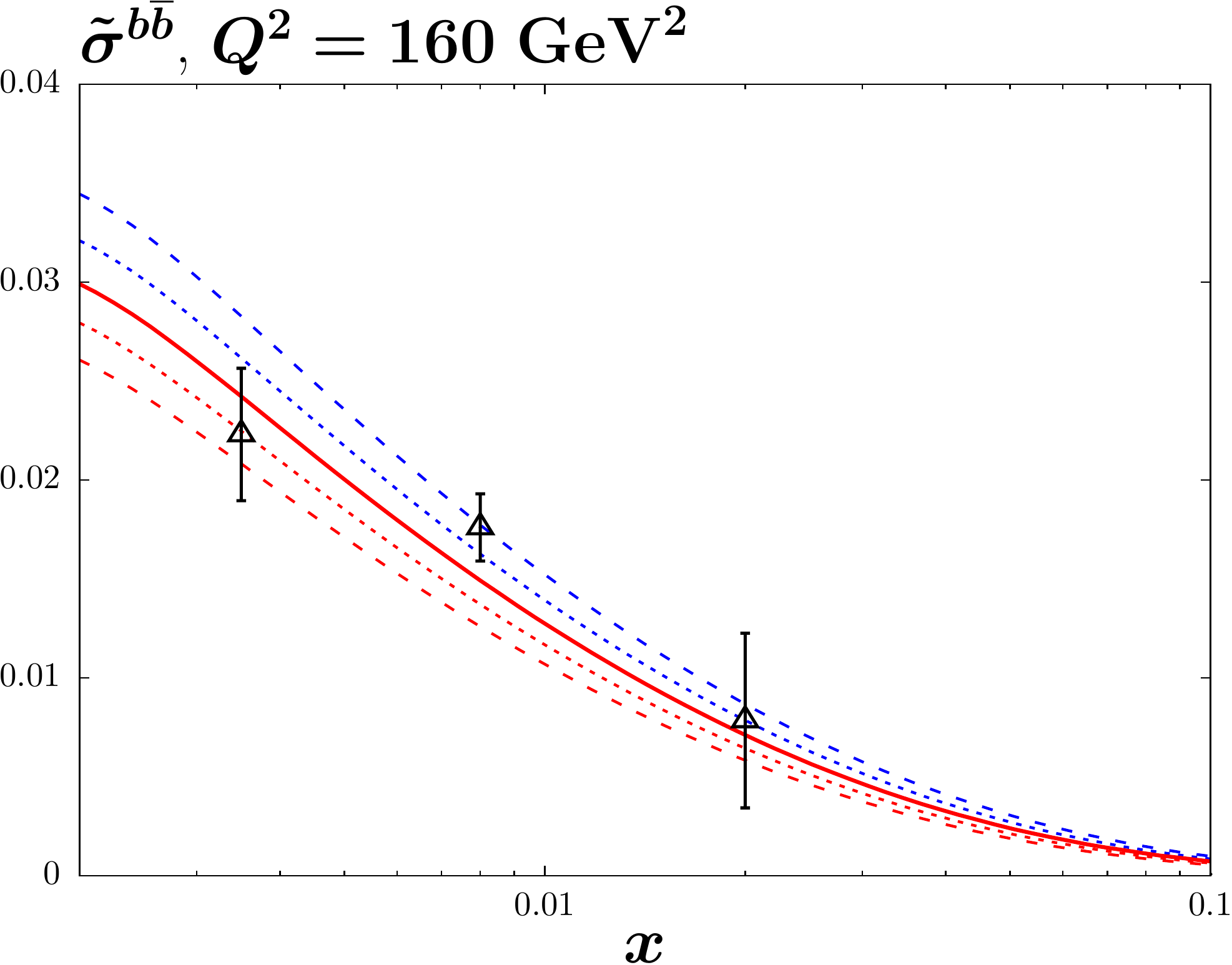}
\includegraphics[scale=0.25]{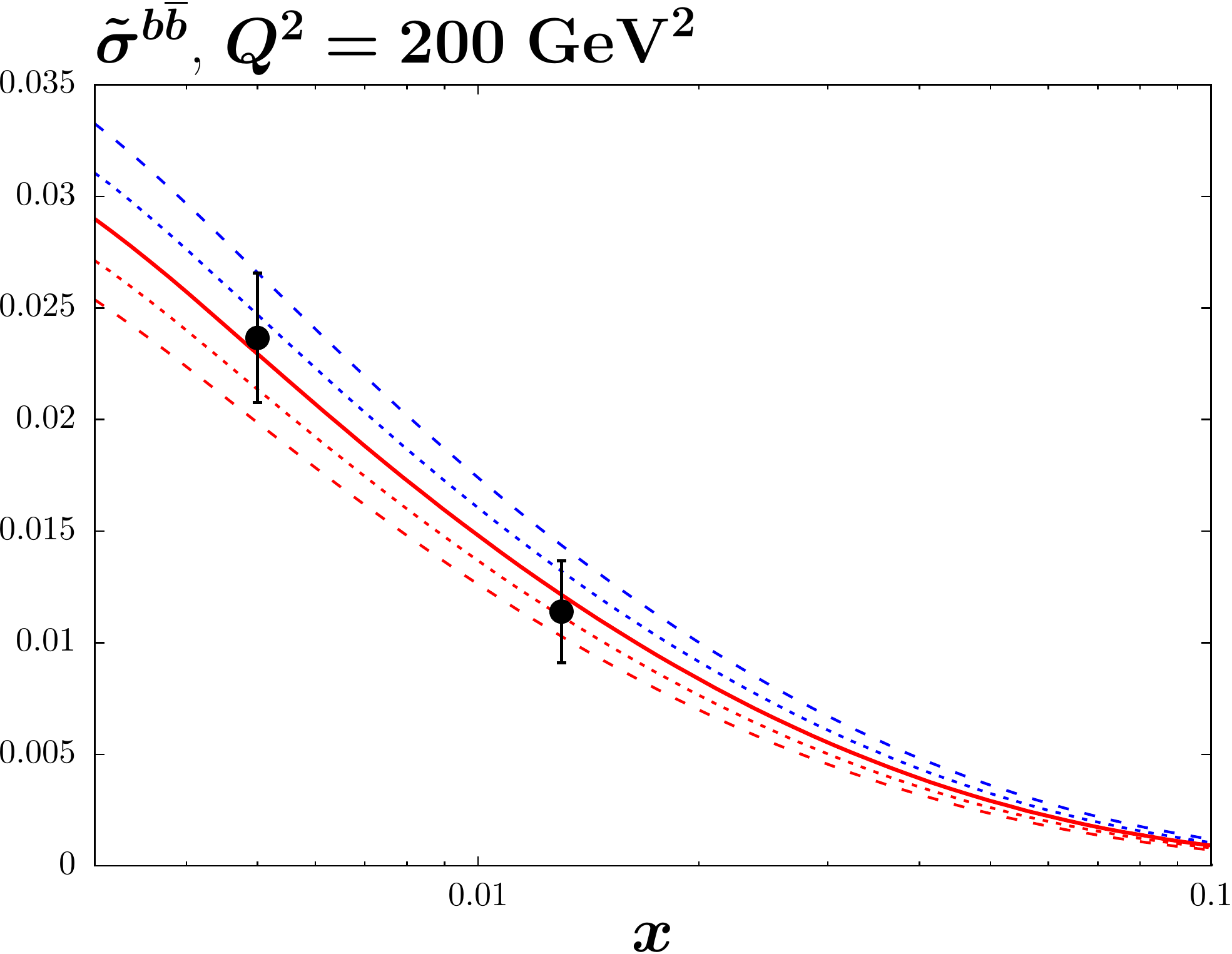}
\includegraphics[scale=0.25]{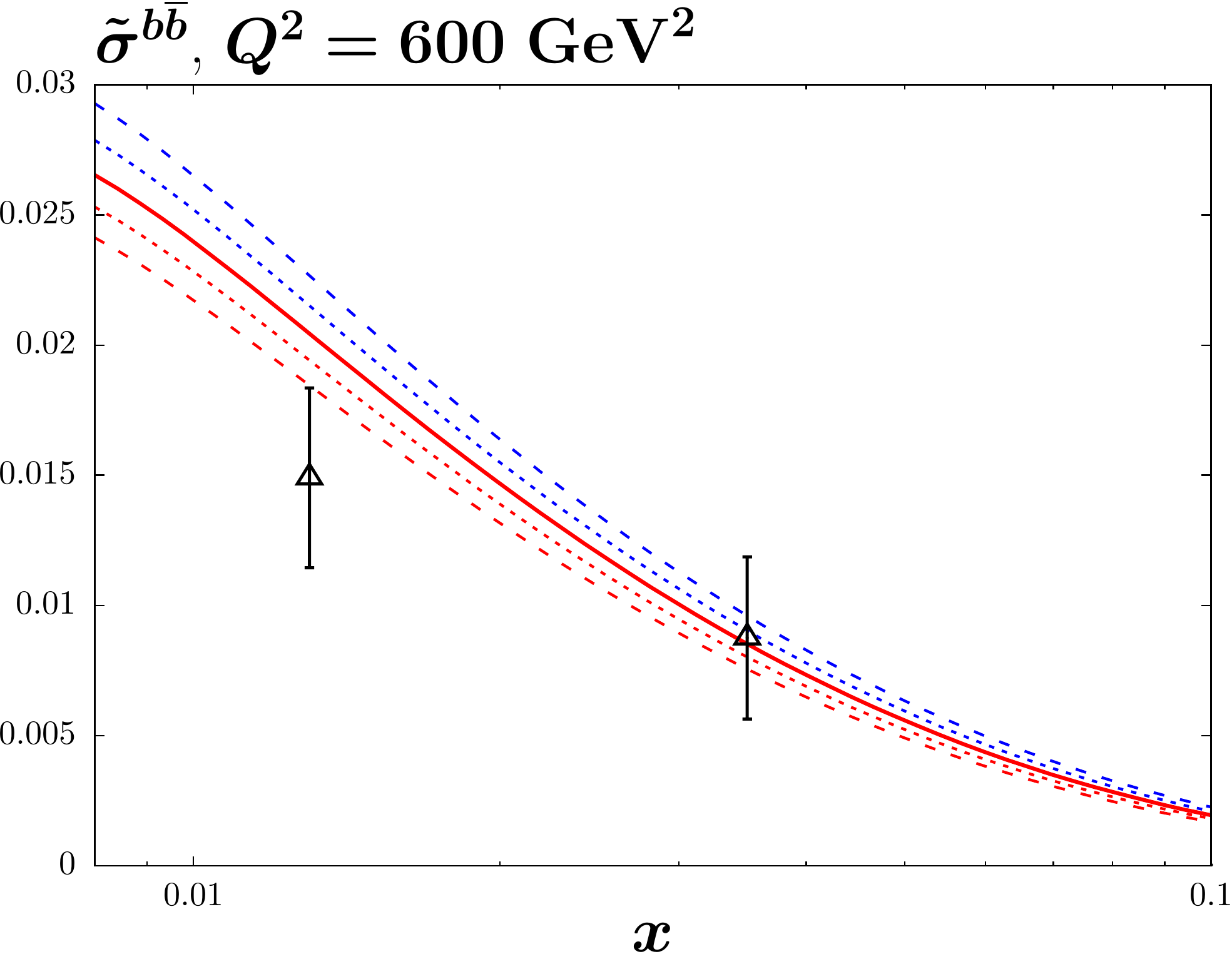}
\includegraphics[scale=0.25]{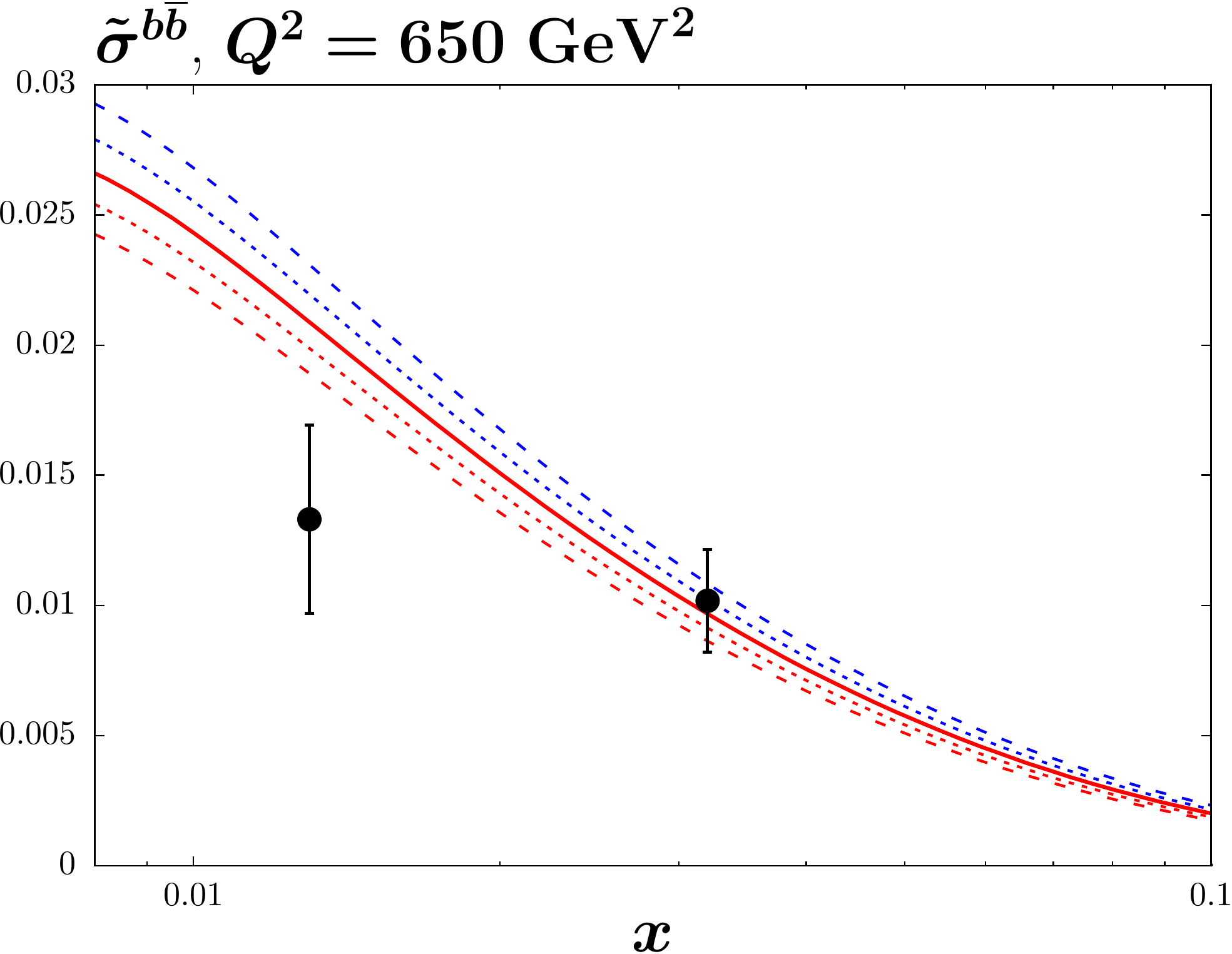}
\includegraphics[scale=0.25]{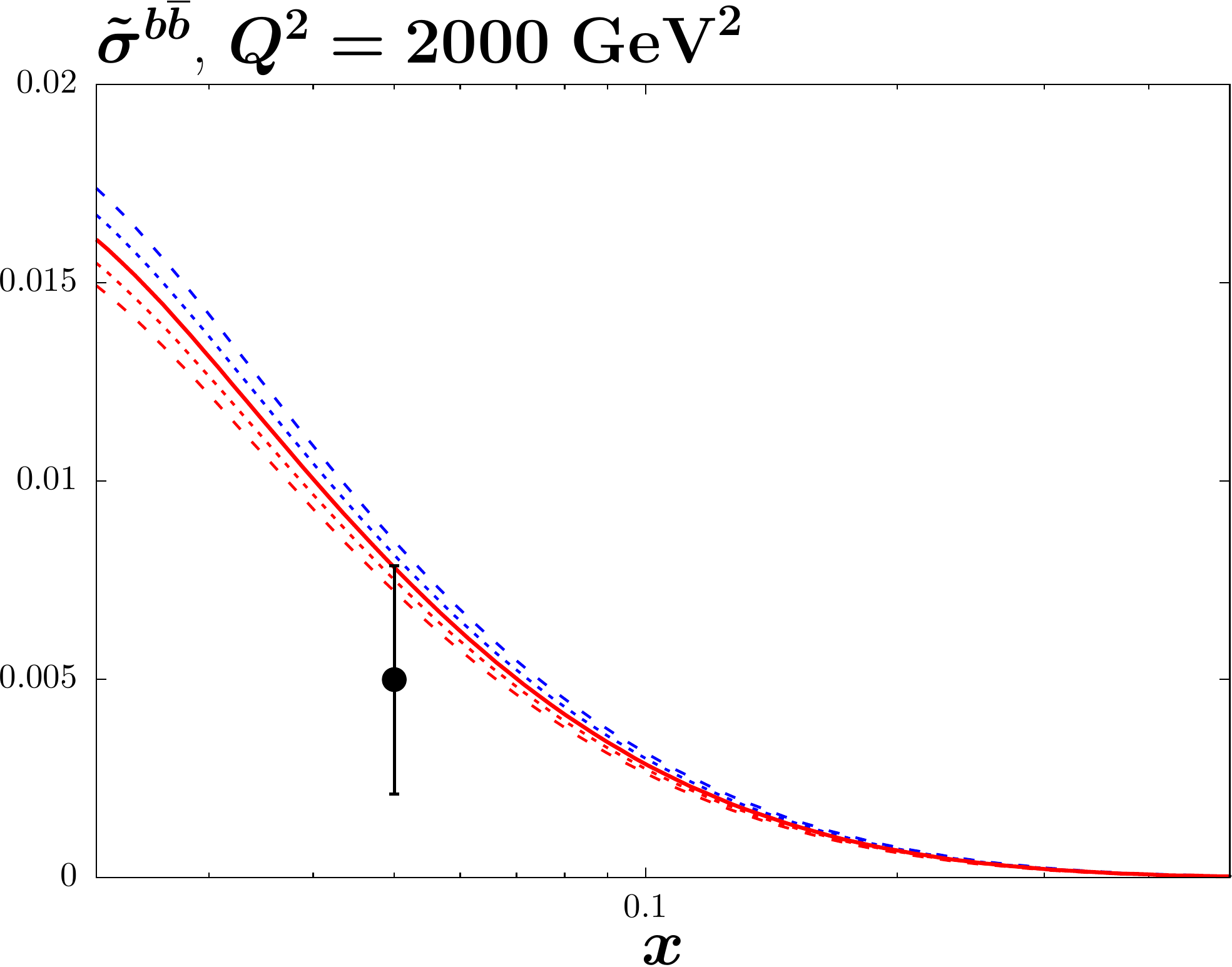}
\caption{\sf  The (unshifted) HERA $\tilde{\sigma}(b{\bar b})$  data versus $x$ at 12 different values of $Q^2$, namely $Q^2=5,~6.5,...2000$ GeV$^2$; the H1~\cite{Aaron:2009af} and ZEUS~\cite{Abramowicz:2014zub} data are shown as solid circular and clear triangular points respectively.  The curves are the NNLO predictions for 5 different values of $m_b$, namely, in descending order, $m_b=4.25,~4.5,~4.75,~5,~5.25$ GeV.}
\label{fig:beautydata}
\end{center}
\end{figure}

In Fig.~\ref{fig:beautydata} the comparison to the (unshifted) HERA beauty data for different values of $m_b$ at NNLO is shown. At low $Q^2$ and for ZEUS data 
in particular, the curves for lower $m_b$ are clearly a better fit to 
unshifted data. However, the low-$m_b^2$ predictions do significantly overshoot
some of the unshifted data points. These predictions will work better with 
the multiplicative definition of uncertainties as the size of the correlated 
uncertainties then scales with the prediction, not the data point (as would 
be the case in the additive definition), or 
equivalently, if data are normalised up to match theory, then so is the 
uncorrelated uncertainty.  

\subsection{Changes in the PDFs}

\begin{figure}
\begin{center}
\vspace*{-1.0cm}
\includegraphics[scale=0.6]{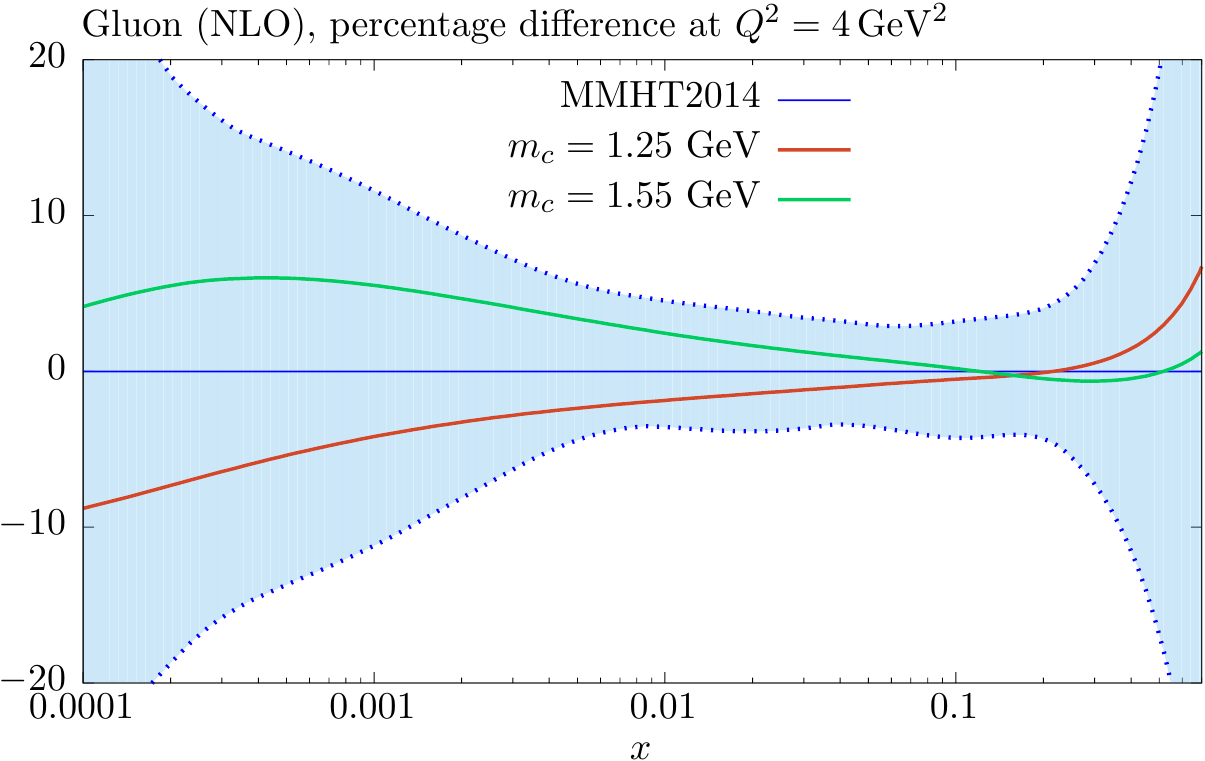}
\includegraphics[scale=0.6]{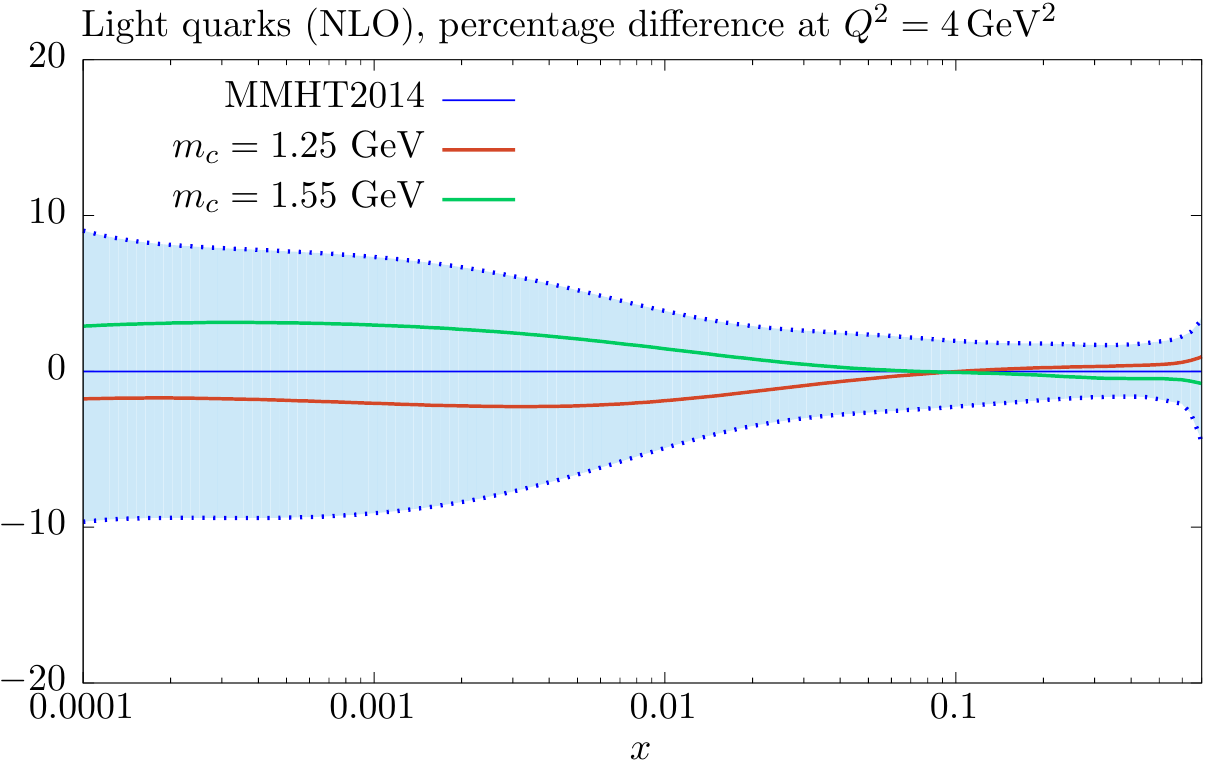}
\includegraphics[scale=0.6]{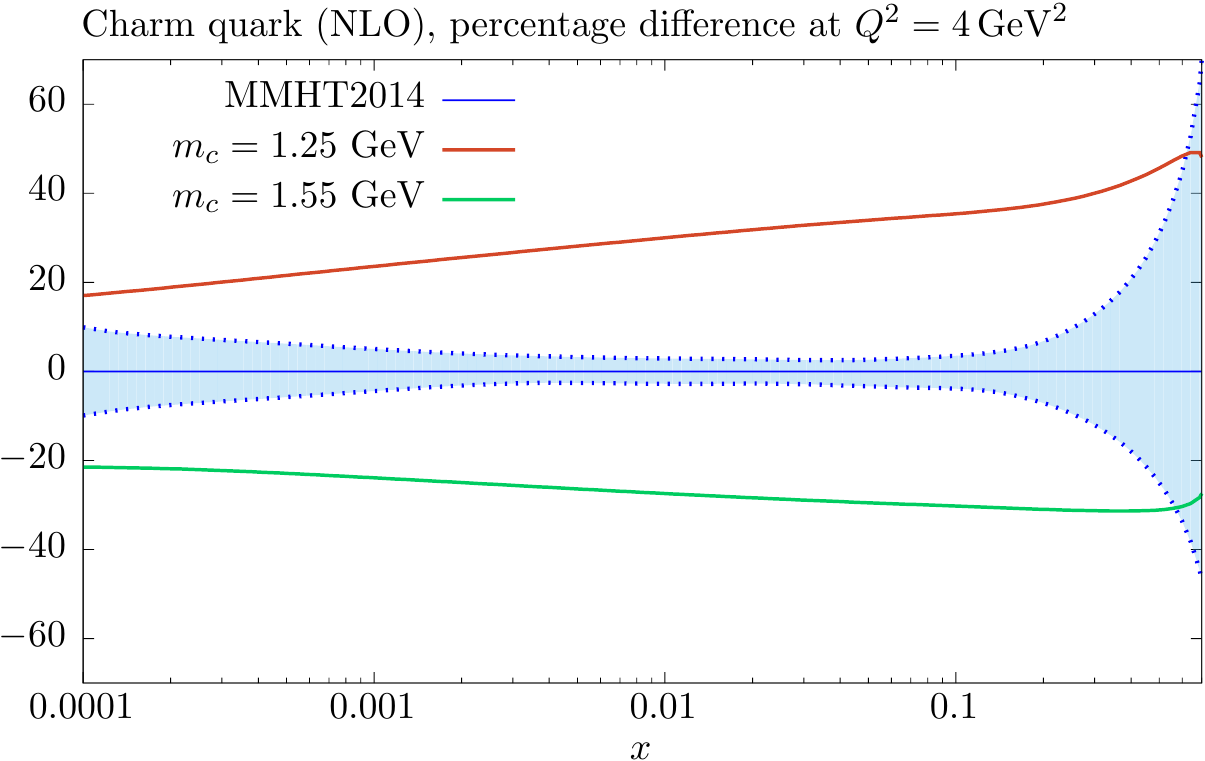}
\caption{\sf The $m_c$ dependence of the gluon, light-quark singlet and charm distributions
at NLO for $Q^2=4~\GeV^2$, compared to the 
standard MMHT2014 distributions with $m_c=1.4~\GeV$ and $m_b=4.75~\GeV$.}
\label{fig:PDFsmcq4}
\end{center}
\end{figure}

\begin{figure}
\begin{center}
\vspace*{-1.0cm}
\includegraphics[scale=0.6]{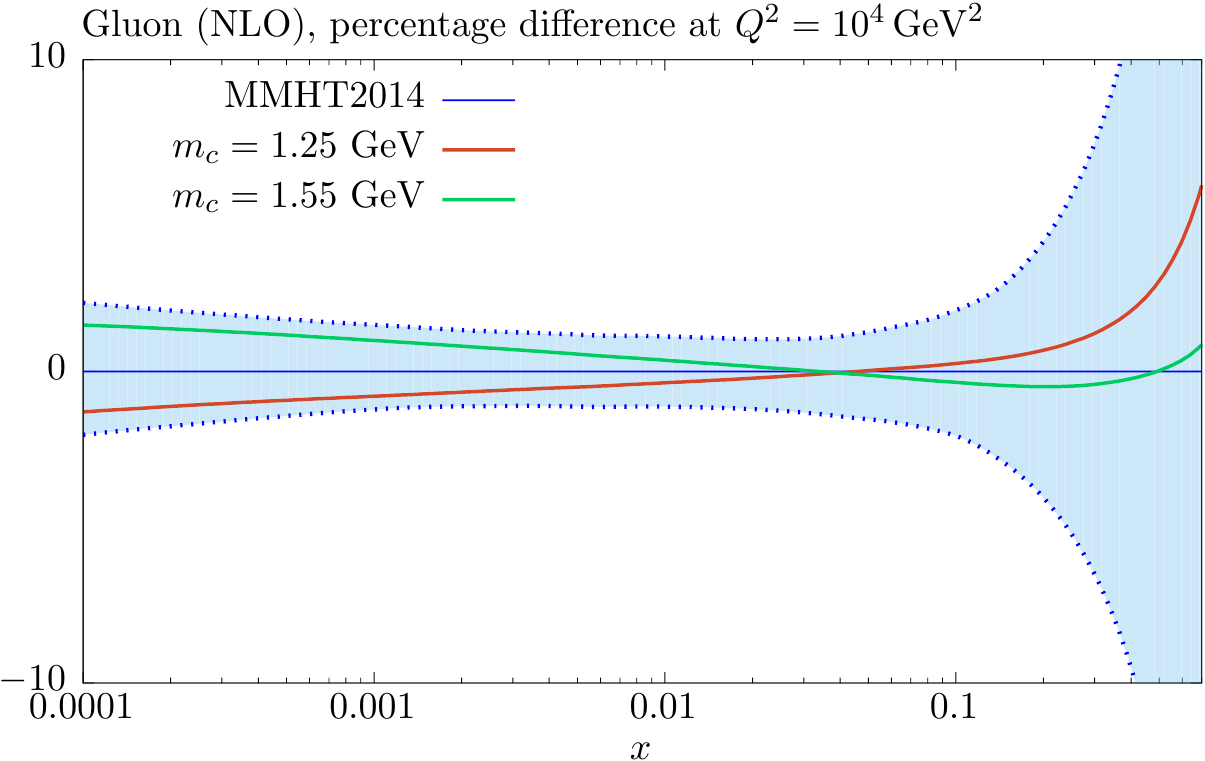}
\includegraphics[scale=0.6]{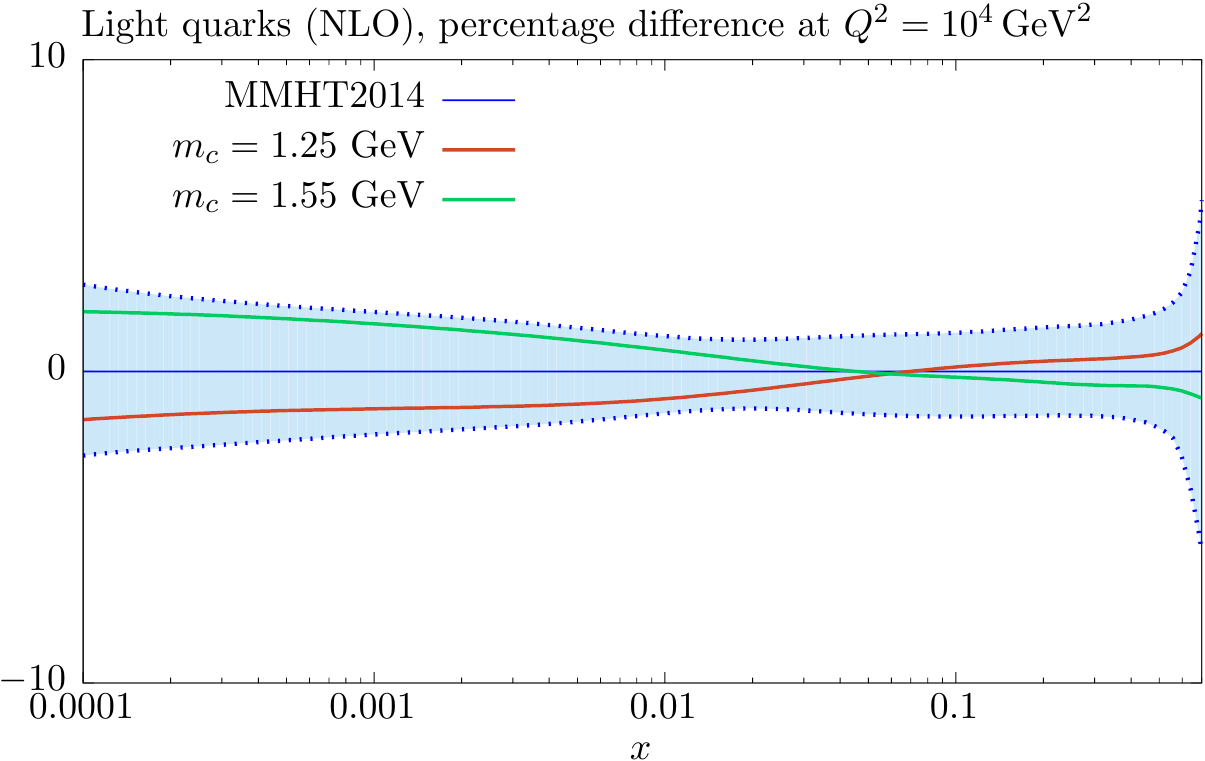}
\includegraphics[scale=0.6]{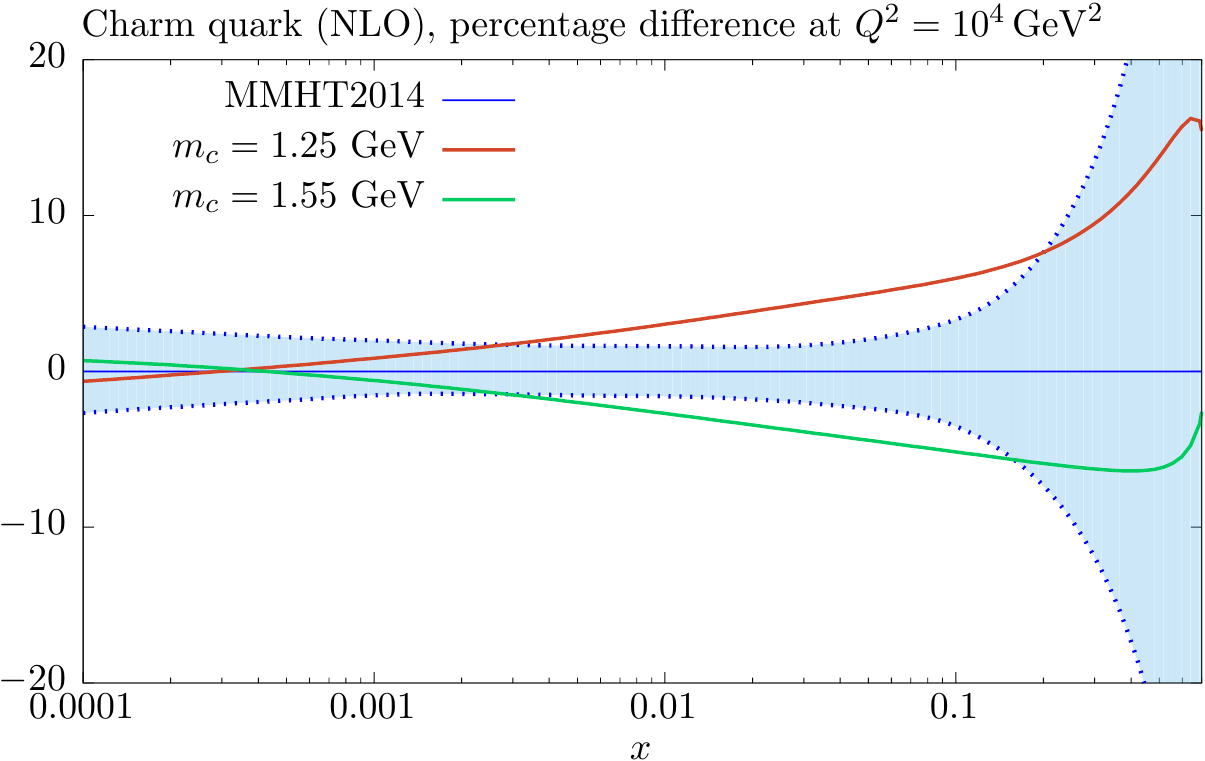}
\caption{\sf The $m_c$ dependence of the gluon, light-quark singlet and charm distributions
at NLO for $Q^2=10^4~\GeV^2$, compared to the 
standard MMHT2014 distributions with $m_c=1.4~\GeV$ and $m_b=4.75~\GeV$.}
\label{fig:PDFsmcq10000}
\end{center}
\end{figure}

We show how the NLO PDFs for $m_c=1.25~\GeV$ and $m_c=1.55~\GeV$ compare to 
the central PDFs in Figs.~\ref{fig:PDFsmcq4} and~\ref{fig:PDFsmcq10000}. Results are very similar at
NNLO, though more complicated to interpret for the charm distribution at low 
$Q^2$ due to the non-zero transition matrix element at $Q^2=m_c^2$ in this 
case. We see at $Q^2=4$ GeV$^2$ (that is, close to the transition point $Q^2=m_c^2$) that the change in the gluon is well within its uncertainty band, though there is a slight increase at smaller $x$ with higher $m_c$ 
(and {\it vice versa}) such that extra gluon quickens the evolution of the 
structure function which is suppressed by larger mass. Similarly the light quark
singlet distribution increases slightly near the transition point for larger 
$m_c$ to make up for the smaller charm contribution to structure functions, and
this is maintained, helped by the increased gluon, at larger scales. In both
cases, however, the changes are within uncertainties for these moderate 
variations in $m_c$. The charm distribution increases at low $Q^2$ for decreasing
$m_c$, and {\it vice versa}, simply due to increased evolution 
length $\ln(Q^2/m_c^2)$. As mentioned before we have identified the transition 
point at which heavy flavour evolution begins with the quark mass. This has 
the advantage that the boundary condition for evolution is zero up to NLO
(with our further assumption that there is no intrinsic charm), though there
is a finite ${\cal O}(\alpha_S^2)$ boundary condition at NNLO in the GM-VFNS, 
available in \cite{Buza:1996wv}. In principle the results on the charm 
distribution at relatively low scales, such as that in Fig.~\ref{fig:PDFsmcq4}
are sensitive to these definitions at finite order, though as the order in QCD 
increases the correction for changes due to different choices of 
transition point arising from the corresponding changes 
in the boundary conditions become smaller and smaller, ambiguities 
always being of higher order than the calculation. 
At scales typical of most of LHC physics, however, the relative change in 
evolution length for the charm distribution is much reduced, as are the 
residual effects of choices relating to choice of transition point and 
intrinsic charm. At these scales the change in the charm distribution 
is of the same general size as the PDF 
uncertainty for fixed $m_c$, as seen in Fig.~\ref{fig:PDFsmcq10000}. 
We also note that the charm structure function at these high scales is 
reasonably well represented by the charm distribution, while at low scales,
certainty including $Q^2=4~\GeV^2$, this is not true. 
Indeed at NNLO the boundary
condition for the charm distribution is negative at very low $x$ if 
the transition point is $m_c^2$, but this is more than compensated for 
by the gluon and light quark initiated cross section. As noted in 
\cite{Thorne}, use of a zero mass scheme becomes unfeasible at NNLO. 
The dependence on the heavy quark cross section at low scales relative to 
the mass is much better gauged from Fig.~\ref{fig:beautydata}.

\begin{figure}
\begin{center}
\vspace*{-1.0cm}
\includegraphics[scale=0.6]{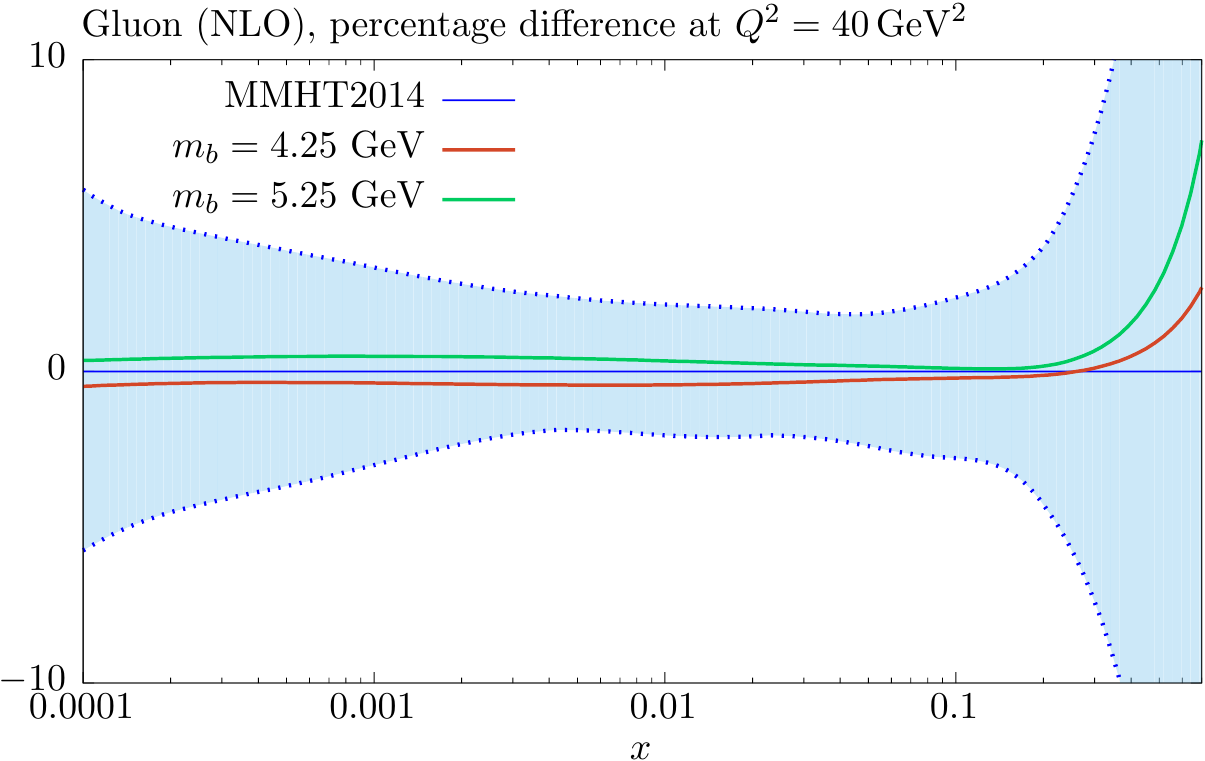}
\includegraphics[scale=0.6]{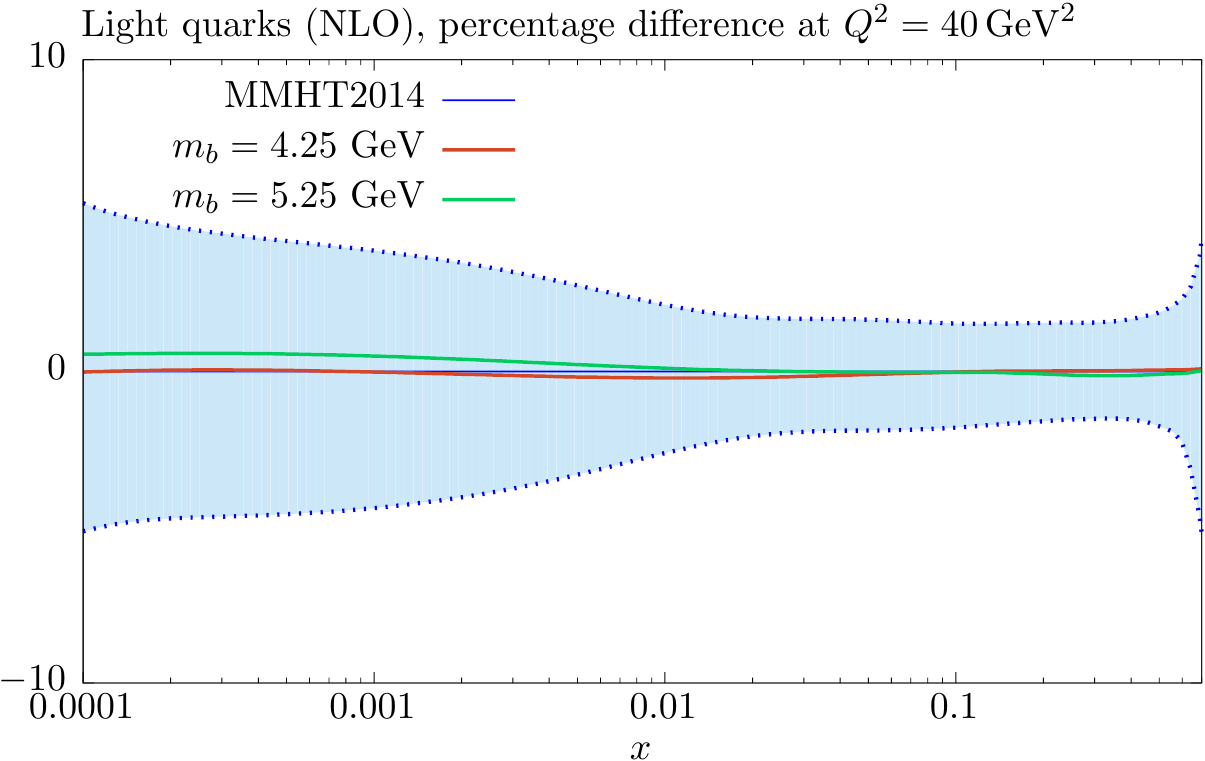}
\includegraphics[scale=0.6]{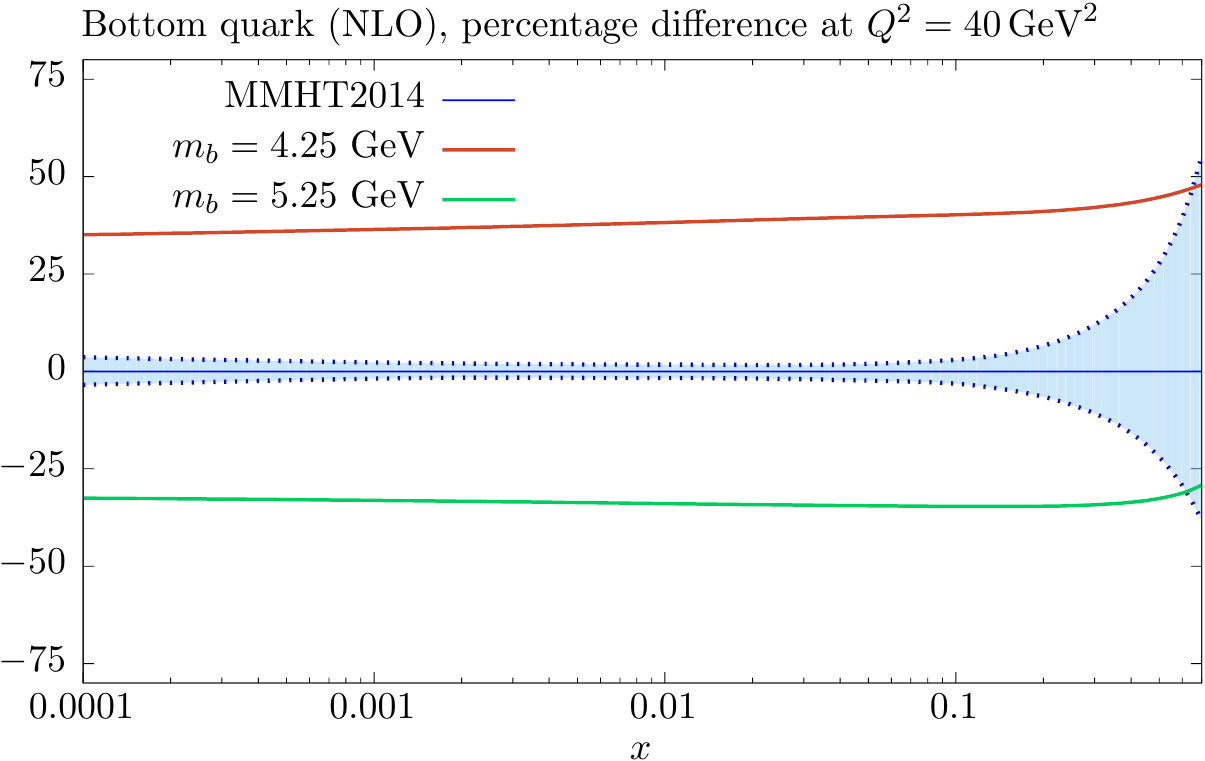}
\caption{\sf The $m_b$ dependence of the gluon, light-quark singlet and charm distributions
at NLO for $Q^2=40~\GeV^2$, compared to the 
standard MMHT2014 distributions with $m_c=1.4~\GeV$ and $m_b=4.75~\GeV$.}
\label{fig:PDFsmbq4}
\end{center}
\end{figure}

\begin{figure}
\begin{center}
\vspace*{-1.0cm}
\includegraphics[scale=0.6]{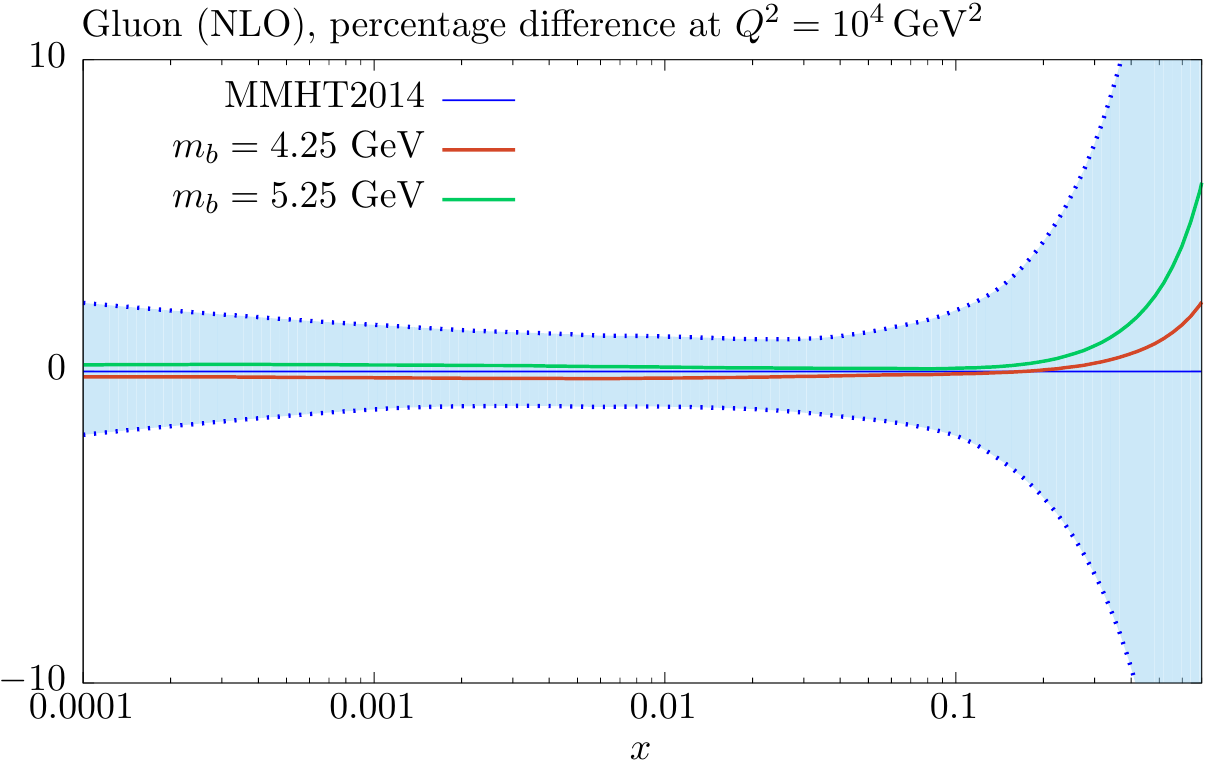}
\includegraphics[scale=0.6]{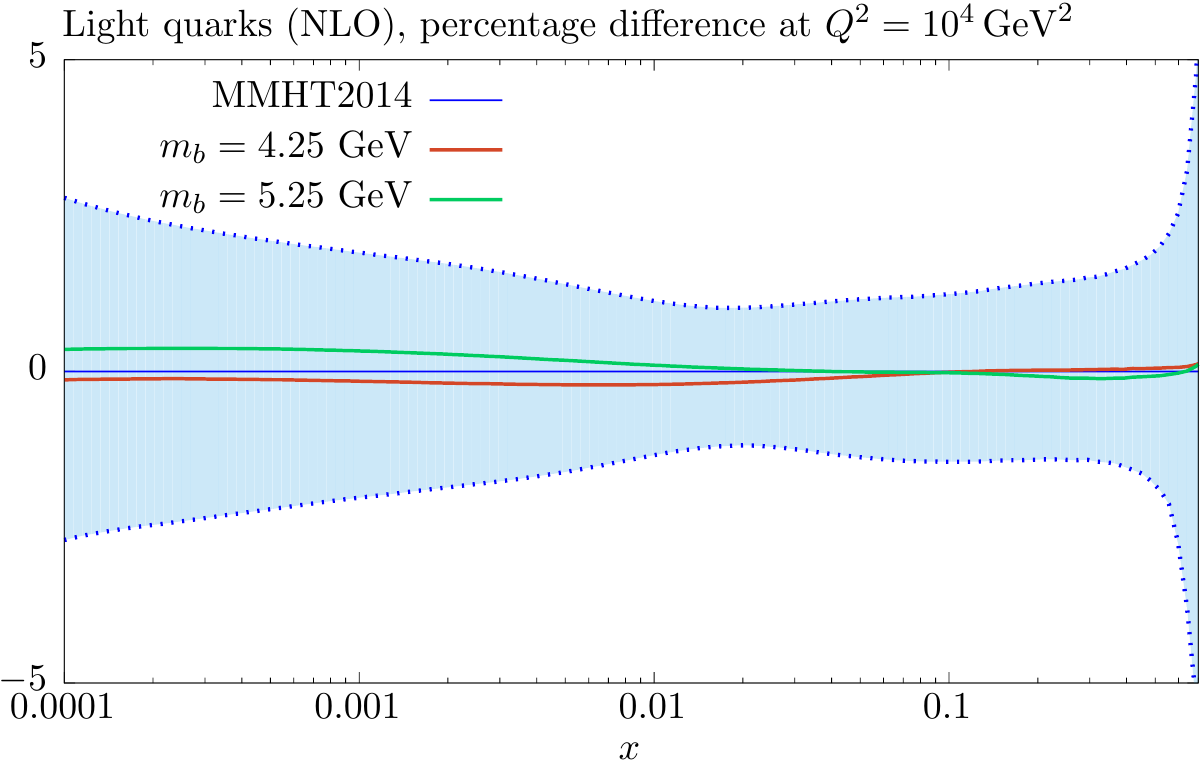}
\includegraphics[scale=0.6]{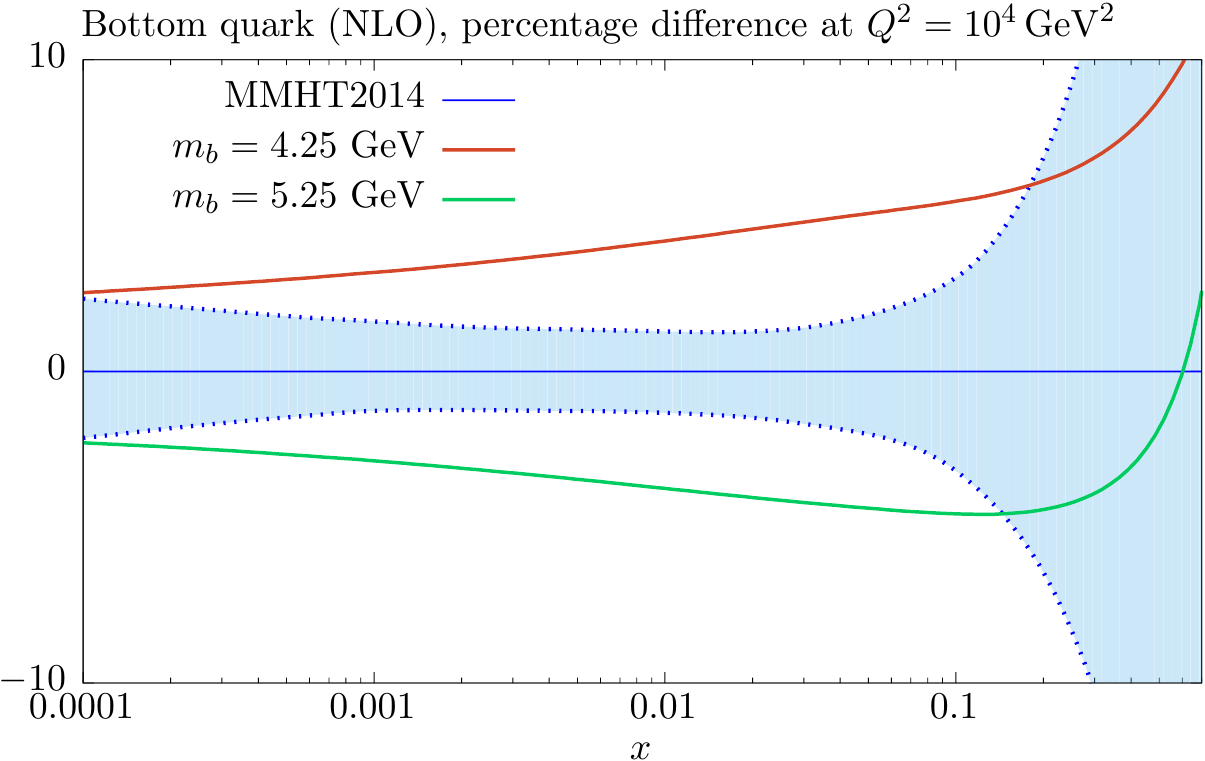}
\caption{\sf The $m_b$ dependence of the gluon, light-quark singlet and charm distributions
at NLO for $Q^2=10^4~\GeV^2$, compared to the 
standard MMHT2014 distributions with $m_c=1.4~\GeV$ and $m_b=4.75~\GeV$.}
\label{fig:PDFsmbq10000}
\end{center}
\end{figure}

The relative changes in the gluon and light quarks for variations in $m_b$ are 
significantly reduced due to the much smaller impact of the beauty contribution 
to the structure functions from the charge-squared weighting, as can be seen 
in Figs.~\ref{fig:PDFsmbq4} and~\ref{fig:PDFsmbq10000}, where we show NLO PDFs for $m_b=4.25~\GeV$ and
$m_b=5.25~\GeV$. At $Q^2=40\, {\GeV^2} \sim 2m_b^2$ the relative change in the 
beauty distribution for a $\sim 10\%$ change in the mass is similar to that 
for the same type of variation for $m_c$. However, the extent to which
this remains at $Q^2=10^4~\GeV^2$ is much greater than the charm case due 
to the smaller evolution length.

\section{Effect on benchmark cross sections \label{sec:cxmcmb}}

\begin{table}[h]
\begin{center}
\renewcommand\arraystretch{1.25}
\begin{tabular}{|l|c|c|c|c|}
\hline
& $\sigma$& PDF unc.& $m_c$ var.& $m_b$ var.  \\
\hline
$\!\! W\,\, {\rm Tevatron}\,\,(1.96~\TeV)$   & 2.78    & ${}^{+0.0017}_{-0.056}$ $\left({}^{+2.0\%}_{-2.0\%}\right)$  & ${}^{+0.0017}_{-0.0086}$  $\left({}^{+0.061\%}_{-0.31\%}\right)$ & ${}^{-0.00092}_{-0.0015}$  $\left({}^{-0.033\%}_{-0.052\%}\right)$       \\   
$\!\! Z \,\,{\rm Tevatron}\,\,(1.96~\TeV)$   & 0.256& ${}^{+0.0052}_{-0.0046}$  $\left({}^{+2.0\%}_{-1.8\%}\right)$ &${}^{+0.00042}_{-0.0011}$ $\left({}^{+0.16\%}_{-0.43\%}\right)$ & ${}^{-0.00029}_{-0.000016}$ $\left({}^{-0.11\%}_{-0.0059\%}\right)$ \\     
\hline                                                                                                                                                                                                                                                   
$\!\! W^+ \,\,{\rm LHC}\,\, (7~\TeV)$        &6.20    & ${}^{+0.103}_{-0.092}$ $\left({}^{+1.7\%}_{-1.5\%}\right)$     &  ${}^{+0.029}_{-0.040}$ $\left({}^{+0.48\%}_{-0.64\%}\right)$  &  ${}^{+0.0043}_{-0.014}$ $\left({}^{+0.070\%}_{-0.22\%}\right)$    \\    
$\!\! W^- \,\,{\rm LHC}\,\, (7~\TeV)$        & 4.31  & ${}^{+0.067}_{-0.076}$ $\left({}^{+1.6\%}_{-1.8\%}\right)$       & ${}^{+0.019}_{-0.022}$ $\left({}^{+0.44\%}_{-0.51\%}\right)$    & ${}^{+0.0059}_{-0.0091}$ $\left({}^{+0.14\%}_{-0.21\%}\right)$ \\    
$\!\! Z \,\,{\rm LHC}\,\, (7~\TeV)$          & 0.964   & ${}^{+0.014}_{-0.013}$ $\left({}^{+1.5\%}_{-1.3\%}\right)$       & ${}^{+0.0074}_{-0.0088}$ $\left({}^{+0.77\%}_{-0.92\%}\right )$  & ${}^{-0.00096}_{-0.00038}$ $\left({}^{-0.10\%}_{-0.039\%}\right)$ \\ \hline    
$\!\! W^+ \,\,{\rm LHC}\,\, (14~\TeV)$       & 12.5      & ${}^{+0.22}_{-0.18}$ $\left({}^{+1.8\%}_{-1.4\%}\right)$     & ${}^{+0.091}_{-0.12}$ $\left({}^{+0.73\%}_{-0.93\%}\right)$     & ${}^{+0.0087}_{-0.037}$ $\left({}^{+0.069\%}_{-0.30\%}\right)$  \\    
$\!\! W^- \,\,{\rm LHC}\,\, (14~\TeV)$       & 9.3     &${}^{+0.15}_{-0.14}$ $\left({}^{+1.6\%}_{-1.5\%}\right)$          & ${}^{+0.064}_{-0.075}$ $\left({}^{+0.69\%}_{-0.81\%}\right)$     & ${}^{+0.012}_{-0.029}$ $\left({}^{+0.13\%}_{-0.31\%}\right)$       \\    
$\!\! Z \,\,{\rm LHC}\,\, (14~\TeV)$         & 2.06   & ${}^{+0.035}_{-0.030}$ $\left({}^{+1.7\%}_{-1.5\%}\right)$         & ${}^{+0.021}_{-0.025}$ $\left({}^{+1.03\%}_{-1.2\%}\right)$      & ${}^{-0.0035}_{-0.0013}$ $\left({}^{-0.17\%}_{-0.062\%}\right)$   \\ 
\hline
    \end{tabular}
\end{center}
\caption{\sf Predictions for $W^\pm$ and $Z$ cross sections (in nb), including leptonic branching, obtained with the NNLO MMHT2014 parton sets. The PDF uncertainties and $m_c$ and 
$m_b$  variations are also shown, where the $m_c$ variation corresponds to $\pm 0.15~\GeV$ and the $m_b$ variation corresponds to $\pm 0.5~\GeV$ , i.e. about
$10\%$ in each case.}
\label{tab:sigmaWZNNLO}   
\end{table}

In this section  we show the variation with $m_c$ and $m_b$ 
for cross sections at the 
Tevatron, and for $7~\TeV$ and $14~\TeV$ at the LHC. Variations for  
$8~\TeV$ and $13~\TeV$ will be very similar to those at $7~\TeV$ and $14~\TeV$
respectively. We calculate the cross sections for $W$ and $Z$ boson, Higgs boson via gluon--gluon fusion and top-quark pair production.
To calculate the cross section we use the same procedure as was used in
\cite{MMHT} and \cite{MMHTas}. That is, for $W, Z$ and Higgs production we use the code provided by W.J. Stirling, based on the
calculation in \cite{WZNNLO}, \cite{HiggsNNLO1} and \cite{HiggsNNLO2}, and for top pair production
we use the procedure and code of \cite{topNNLO}. Here our primary aim is not to present definitive predictions or to compare in detail to other PDF sets, as both these results are frequently provided in the literature with very specific choices of codes, scales and parameters which may differ from those used here. Rather, our main objective is to illustrate the relative influence of varying $m_c$ and $m_b$ for these benchmark processes.

\begin{table}[h]
\begin{center}
\renewcommand\arraystretch{1.25}
\vspace{0.5cm}
\begin{tabular}{|l|c|c|c|c|}
\hline
& $\sigma$& PDF unc.& $m_c$ var.& $m_b$ var. \\
\hline
$t\overline{t}$ $ {\rm Tevatron}\,\,(1.96~\TeV)$   & 7.5    & ${}^{+0.21}_{-0.20}$ $\left({}^{+2.8\%}_{-2.7\%}\right)$ & ${}_{+0.077}^{-0.059}$  $\left({}_{+1.0\%}^{-0.78\%}\right)$  & ${}_{+0.0015}^{+0.0088}$  $\left({}_{+0.20\%}^{+0.12\%}\right)$   \\   
$t\overline{t}$  ${\rm LHC}\,\, (7~\TeV)$              &176    & ${}^{+3.9}_{-5.5}$ $\left({}^{+2.2\%}_{-3.1\%}\right)$   &  ${}_{+1.4}^{-1.1}$ $\left({}_{+0.77\%}^{-0.60\%}\right)$   &  ${}_{-0.009}^{+0.77}$ $\left({}_{-0.0051\%}^{+0.44\%}\right)$        \\    
$t\overline{t}$ ${\rm LHC}\,\, (14~\TeV)$             & 970   &${}^{+16}_{-20}$ $\left({}^{+1.6\%}_{-2.1\%}\right)$      & ${}_{+3.1}^{-3.0}$ $\left({}_{+0.32\%}^{-0.31\%}\right)$       & ${}_{-1.7}^{+3.1}$ $\left({}_{+0.17\%}^{-0.32\%}\right)$      \\
\hline
    \end{tabular}
\end{center}
\caption{\sf Predictions for $t\overline{t}$ cross sections (in nb), obtained with the NNLO MMHT2014 parton sets. The PDF uncertainties and $m_c$ and 
$m_b$  variations are also shown, where the $m_c$ variation corresponds to $\pm 0.15~\GeV$ and the $m_b$ variation corresponds to $\pm 0.5~\GeV$. }
\label{tab:sigmatNNLO}   
\end{table}

\begin{table}[h]
\begin{center}
\renewcommand\arraystretch{1.25}
\begin{tabular}{|l|c|c|c|c|}
\hline
& $\sigma$& PDF unc.& $m_c$ var.& $m_b$ var.\\
\hline
Higgs $ {\rm Tevatron}\,\,(1.96~\TeV)$   & 0.87    & ${}^{+0.024}_{-0.030}$ $\left({}^{+2.7\%}_{-3.4\%}\right)$ &  ${}^{-0.0060}_{+0.0070}$  $\left({}^{-0.68\%}_{+0.79\%}\right)$  & ${}^{+0.0042}_{-0.0011}$  $\left({}^{+0.48\%}_{-0.13\%}\right)$  \\       
Higgs  ${\rm LHC}\,\, (7~\TeV)$              &14.6    & ${}^{+0.21}_{-0.29}$ $\left({}^{+1.4\%}_{-2.0\%}\right)$     & ${}^{+0.025}_{-0.019}$ $\left({}^{+0.17\%}_{-0.13\%}\right)$   &  ${}^{+0.049}_{-0.044}$ $\left({}^{+0.34\%}_{-0.30\%}\right)$           \\    
Higgs ${\rm LHC}\,\, (14~\TeV)$             & 47.7   &${}^{+0.63}_{-0.88}$ $\left({}^{+1.3\%}_{-1.8\%}\right)$      & ${}^{+0.27}_{-0.22}$ $\left({}^{+0.57\%}_{-0.48\%}\right)$       & ${}^{+0.16}_{-0.16}$ $\left({}^{+0.34\%}_{-0.33\%}\right)$            \\    
\hline
    \end{tabular}
\end{center}
\caption{\sf Predictions for the Higgs boson cross sections (in nb), obtained with the NNLO MMHT 2014 parton sets. The PDF uncertainties and $m_c$ and 
$m_b$  variations are also shown, where the $m_c$ variation corresponds to $\pm 0.15~\GeV$ and the $m_b$ variation corresponds to $\pm 0.5~\GeV$.}
\label{tab:sigmahNNLO}   
\end{table}

We show the predictions for the default MMHT2014 PDFs, with PDF uncertainties, 
and the relative changes due to changing $m_c$ from $1.25~\GeV$ to $1.55~\GeV$, and $m_b$ from $4.25~\GeV$ to $5.25~\GeV$, i.e. changing the default values by 
approximately $10\%$ in each case. The dependence of the benchmark predictions on the value of $m_c$ in Tables 3-5 reflects the behaviour of the gluon with $\sqrt{s}$.
The changes in cross section generally scale linearly in variation of masses away from the default values to a good approximation, although for $m_b$, where the cross section sensitivity to the mass choice is often small, this is less true, and in some cases the cross section is even found to decrease or increase in both directions away from the best fit mass.

We begin with the predictions for the $W$ and $Z$ production cross sections. 
The results at NNLO are shown in Table \ref{tab:sigmaWZNNLO}.
The PDF uncertainties on the cross 
sections are $2\%$ at the Tevatron and slightly smaller at the LHC -- the lower beam energy at the
Tevatron meaning the cross sections have more contribution from higher $x$ where the PDF uncertainties
increase. The $m_c$ variation is at most about $0.4\%$ at the Tevatron and is 
$0.5-1\%$ at 
the LHC, being larger at 14~TeV.   The results at NLO are very similar.

In Table \ref{tab:sigmatNNLO} we show the analogous results for the top-quark pair  production cross section. 
At the Tevatron the PDFs are probed in the region $x\approx 0.4/1.96\approx 0.2$, and the main production is from the $q{\bar q}$ channel.  
At the LHC the dominant production at higher energies (and with a proton-proton rather than 
proton-antiproton collider) is gluon-gluon fusion, with the central $x$ value probed being
$x\approx 0.4/7 \approx 0.06$ at 7~TeV, and $x\approx 0.4/14\approx 0.03$ at 14~TeV. The PDF uncertainties on the cross 
sections are nearly $3\%$ at the Tevatron, similar for 7~TeV at the LHC, but a little smaller at 14 TeV as there is less sensitivity to the high-$x$ gluon. The $m_c$ variation are slightly less than $1\%$ at the Tevatron and for 7~TeV at the LHC, but rather lower at 14~TeV since the $x$ probed is near the fixed point for the gluon (see Fig.~\ref{fig:PDFsmcq10000}).

In Table \ref{tab:sigmahNNLO} we show the uncertainties in the rate of Higgs boson production from gluon-gluon fusion. At the 
Tevatron the dominant $x$ range probed, i.e. $x\approx 0.125/1.96 \approx 0.06$, corresponds to a region where
the gluon distribution falls as $m_c$ increases and at the LHC where $x \approx 0.01-0.02$ at central rapidity the gluon increases as $m_c$ increases, though 
at 7~TeV we are only just below the fixed point. At the Tevatron the
resultant uncertainty is $\sim 0.7\%$. At the LHC at 7~TeV it is in the opposite
direction but only $\sim 0.1\%$, whereas at 14~TeV it has increased to near
$0.5\%$. 

As in \cite{MSTWhf} we recommend that in order to estimate the total uncertainty due to PDFs and the quark masses
it is best to add the variation due to the variation in quark mass in quadrature with the PDF uncertainty, or the PDF+$\alpha_S$ uncertainty, if 
the $\alpha_S$ uncertainty is also used.

\section{PDFs in three- and four-flavour-number-schemes}

In our default studies we work in a general-mass variable-flavour-number-scheme (GM-VFNS) with a maximum of 5 active flavours. This means that we start
at our input scale of $Q_0^2=1~\GeV^2$ with three active light flavours. 
At the transition point $m_c^2$ the charm quark starts evolution and then 
at $m_b^2$ the beauty quark also starts evolution. The evolution is in terms 
of massless splitting functions, and at high $Q^2$ the contribution from 
charm and bottom quarks lose all mass dependence other than that in the 
boundary conditions at the chosen transition point. The explicit mass 
dependence is included at lower scales, but falls away like inverse powers 
as $Q^2/m^2_{c,b} \to \infty$. We do not currently ever consider the top quark as 
a parton. 

We could alternatively keep the information about the heavy quarks only
in the coefficient functions, i.e. the heavy quarks would only be 
generated in the final state. This is called a fixed-flavour-number-scheme 
(FFNS). One example would be where neither charm and beauty exist as partons.
This would be a 3-flavour FFNS. An alternative would be to let charm 
evolution turn on but never allow beauty to be treated as a parton. This is 
often called a 4-flavour FFNS. We will use this notation, but strictly speaking it 
is a GM-VFNS with a maximum of 4 active flavours.

One might produce the partons for the 3- and 4-flavour FFNS by performing 
global fits in these schemes. However, it was argued in  
\cite{Thorne:2008xf} that the fit to structure function data is not optimal 
in these schemes. Indeed, evidence for this has been provided in 
\cite{Thorne,ThorneFFNS,NNPDFgmvfns}. Moreover, 
much of the data (for example, on inclusive jets and $W,Z$ production
at hadron colliders) is not known to NNLO in these schemes, and is 
very largely at scales where $m_{c,b}$ are relatively very small. So it is clear that the 
GM-VFNS are more appropriate. Hence, in \cite{Martin:2006qz} it was decided
to make available PDFs in the 3- and 4-flavour schemes simply by using the input 
PDFs obtained in the GM-VFNS, but with evolution of the beauty quark, or both 
the beauty and charm quark, turned off. This procedure was continued in \cite{MSTWhf} and is 
the common choice for PDF groups who fit using a GM-VFNS but make 
PDFs available with a maximum of 3- or 4- active flavours. Hence, here, we continue 
to make this choice for the MMHT2014 PDFs. 

We make PDFs available with a maximum of 3 or 4 active flavours for 
the NLO central PDFs and their uncertainty eigenvectors for both 
the standard choices of $\alpha^{n_{f,\max}=5}_S(M_Z^2)$ of 0.118 and 0.120, and for the 
NNLO central PDF and the uncertainty eigenvectors for  
the standard choice of $\alpha^{n_{f,\max}=5}_S(M_Z^2)$ of 0.118. We also provide PDF 
sets with $\alpha_S(M_Z^2)$ displaced by 0.001 from these default values, 
so as to assist with the calculation of $\alpha_s$ uncertainties in the 
different flavour schemes. Finally, we make available PDF sets with different 
values of $m_c$ and $m_b$ in the different fixed-flavour schemes. 

By default, when the charm or beauty quark evolution is turned off, we
also turn off the contribution of the same quark to the running coupling. 
This is because most calculations use this convention when these quarks are entirely 
final state particles. This results in the coupling 
running more quickly. So if the coupling at $Q_0^2$ is chosen so that 
$\alpha^{n_{f,\max}=5}_S(M_Z^2)\approx 0.118$, then we find that $\alpha^{n_{f,\max}=3}_S(M_Z^2)\approx 0.105$
and $\alpha^{n_{f,\max}=4}_S(M_Z^2)\approx 0.113$. There are sometimes 
cases where a set of PDFs with no beauty quark but with 5-flavour running 
 coupling is desired, e.g. \cite{Cascioli:2013era}. After the publication
of \cite{MSTWhf}, PDF sets with this definition were made available. Here 
we make available PDFs for the central sets together with their eigenvectors with 
a maximum of 4 active flavours, but the beauty quark included in 
the running of the coupling. This type of PDF has also been considered very recently in~\cite{Bertone:2015gba}.  
   
\begin{figure} [h]
\begin{center}
\vspace*{-1.0cm}
\includegraphics[scale=0.65]{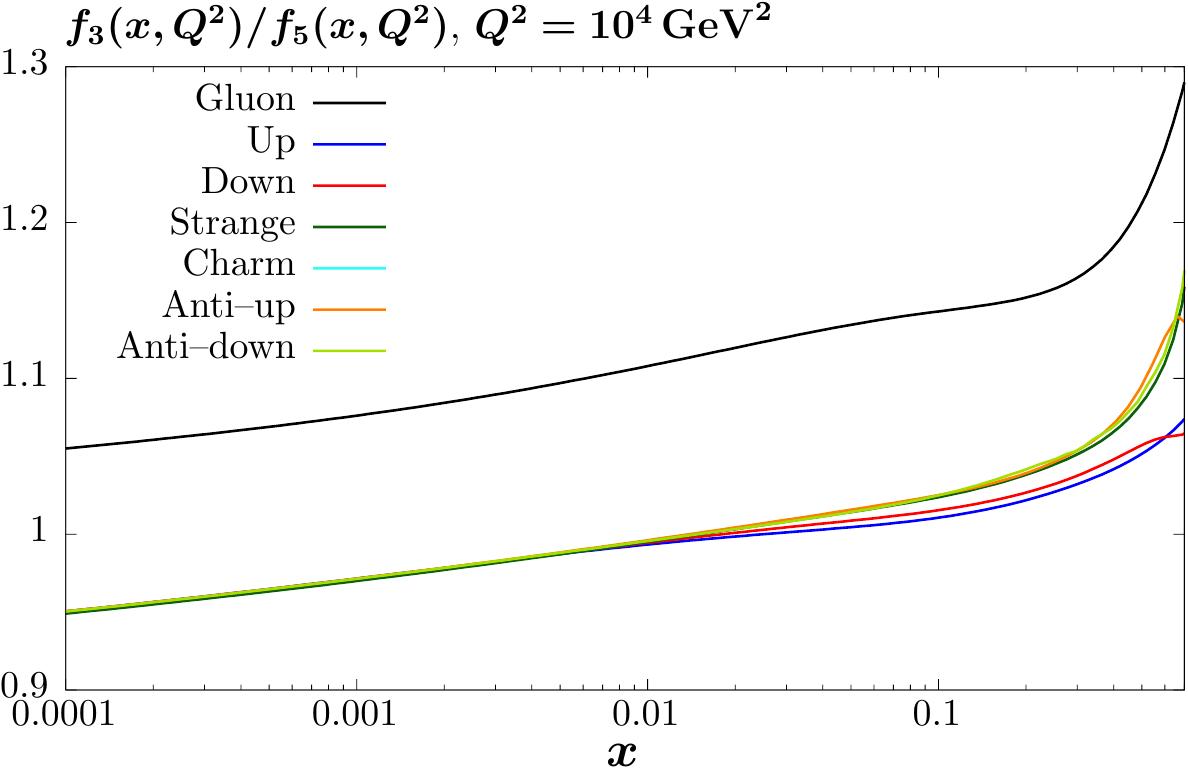}
\includegraphics[scale=0.65]{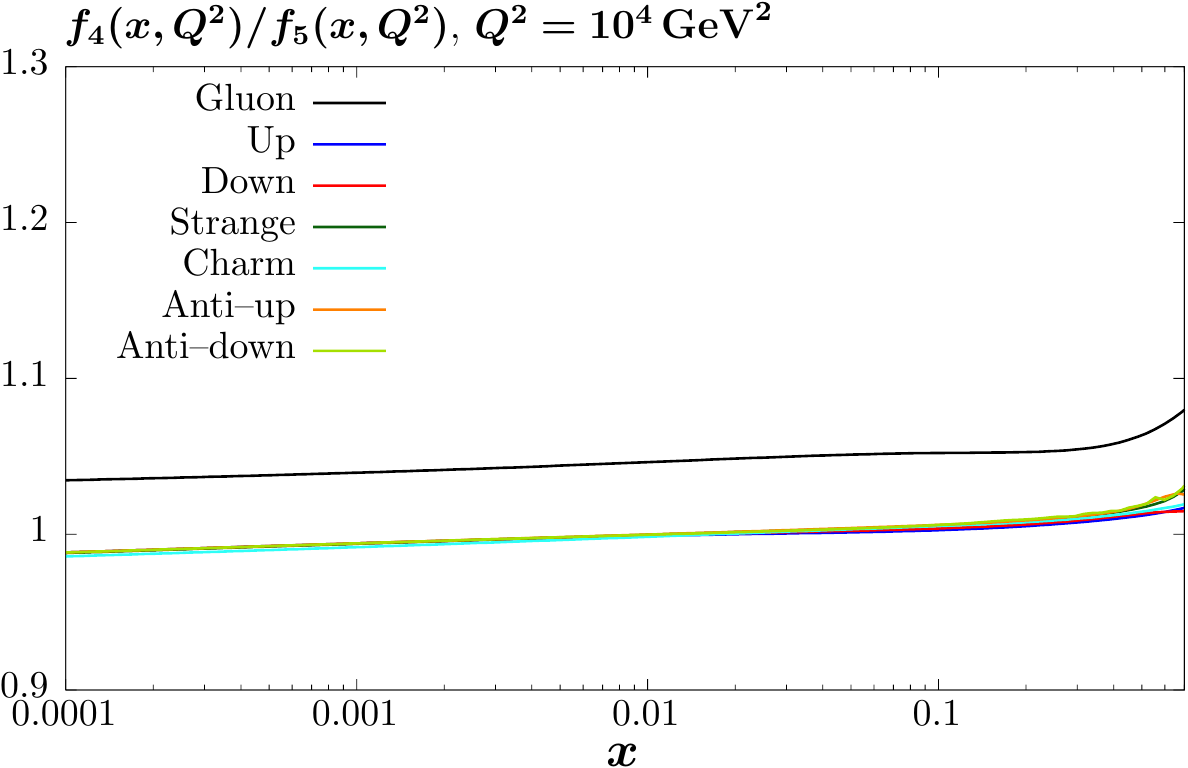}
\includegraphics[scale=0.65]{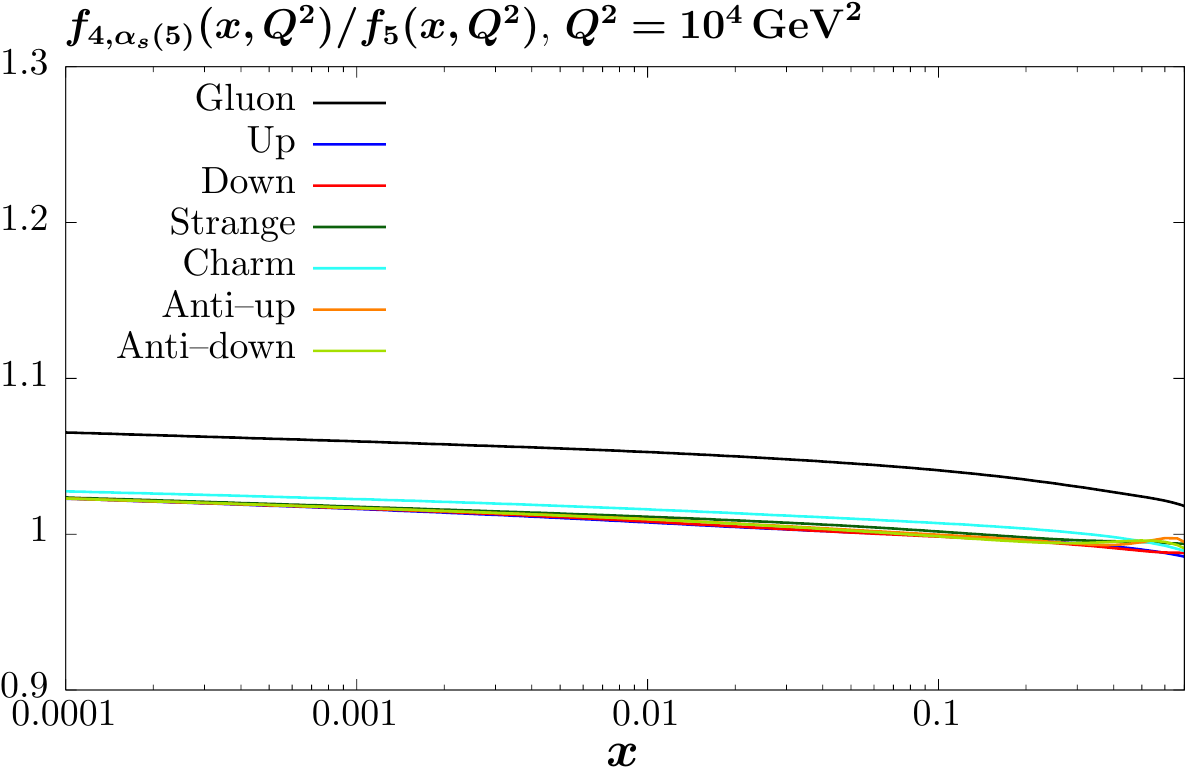}
\caption{\sf The ratio of the different fixed flavour PDFs to the standard 5 flavour PDFs at NNLO and 
at $Q^2=10^4~\GeV^2$. The 3 and 4 flavour schemes are show in the top left and right plots, while the 4 flavour scheme with 5 flavours in the running of $\alpha_S$ is shown in the bottom plot.}
\label{fig:PDFs3and4}
\end{center}
\end{figure}  

The variation of the PDFs defined with a maximum number of 3 and 4 flavours, 
compared to our default of 5 flavours, is shown at $Q^2=10^4~\GeV^2$ in 
Fig. \ref{fig:PDFs3and4} for NNLO PDFs. The general form of the differences 
are discussed in detail in Section 4 of \cite{MSTWhf}, and are primarily due to 
two effects. For fewer active quarks there is less gluon branching, so the 
gluon is larger if the flavour number is smaller. Also, as $Q^2$ increases the 
coupling gets smaller for fewer active quarks, so evolution is generally 
slower, which means partons decrease less quickly for large $x$ and 
grow less quickly at small $x$. The latter effect dominates for quark evolution, 
while for the gluon the two effects compete at small $x$. For the case 
where the maximum number of flavours is 4, but the coupling has five-flavour 
evolution, the overwhelming effect is that the gluon is larger --- effectively
replacing the missing beauty quarks in the momentum sum rule. However, the 
increase in the gluon is maximal at small $x$, where the increased coupling 
compared to the case where we use the 4-flavour coupling leading to increased 
loss of gluons at at high $x$ from evolution.

\section{Renormalization Schemes}

At present most PDF fitting groups, including the most recent updates
\cite{MMHT,NNPDF3,CT14,HERA2}, use the pole mass definition for the heavy 
quarks. Hence, we have remained with this definition in our investigation of 
quark mass dependence in this article. 
The analyses in \cite{ABM14,JR14} use the $\MS$ definition, following 
the developments in \cite{Msbarcharm}. 
The latter analyses perform their fits in the fixed-flavour-number-scheme
(FFNS), 
while all the others groups use a general-mass variable-flavour-number-scheme.
There is no fundamental obstacle to switching between the two renormalization 
schemes using either approach. The mass-dependence in a GM-VFNS appears 
in entirety from the FFNS
coefficient functions and in the transition matrix elements which 
set the boundary conditions for the (massless) evolution of the charm and
beauty quarks. These are used along with the FFNS coefficient
functions to define the GM-VFNS coefficient functions which tend to 
the massless versions as $m_{c,b}^2/Q^2 \to 0$.  Under a change in definition
of the quark mass\footnote{Note that $d^1(\mu_R^2)=4/3\pi$ if $\mu_R=m$.}
\be
m^{\rm pole} =  m(\mu_R)(1+\alpha_S(\mu_R^2)d^1(\mu_R^2) + \cdots)
\label{eq:massdef}
\ee
the coefficient functions and transition matrix elements can be 
transformed from one mass scheme to the other straightforwardly, as 
illustrated in eq.(8) of \cite{Msbarcharm}, and the mass in GM-VFNS defined in the $\MS$ renormalization scheme. 

However, there is more sensitivity to the definition of the mass in a 
FFNS at given order than in a GM-VFNS. At LO there is 
no mass scheme dependence in the same way that there is no renormalisation 
scheme dependence of any sort. At NLO in the FFNS the variation of 
the LO ${\cal O}(\alpha_S)$ coefficient function under the change in 
eq.(\ref{eq:massdef}) leads to a change in the NLO ${\cal O}(\alpha_S)$
coefficient function. Some NLO GM-VFNS definitions (e.g. the SACOT($\chi$)
\cite{Tung} and the FONLL-A \cite{Forte:2010ta}) only use the 
FFNS coefficient functions at ${\cal O}(\alpha_S)$. The transition matrix 
element for heavy-quark evolution in an NLO GM-VFNS is also defined
at ${\cal O}(\alpha_S)$ (and indeed is zero with the standard choice 
$\mu_F=m_{c,b}$), so neither depend on the mass definition, and the NLO
GM-VFNS is independent of the mass scheme \cite{Bertone:2012wi}. 

Some 
NLO GM-VFNS definitions do use the ${\cal O}(\alpha_S^2)$ FFNS coefficient
functions. Hence, these will contain some dependence on the mass scheme. 
However, in the original TR \cite{TR} and then the TR' \cite{TR1} schemes
this contribution is frozen at $Q^2=m_{c,b}^2$, so becomes relatively very small
at high $Q^2$. In the ``optimal'' TR' scheme \cite{Thorne}, and in the 
FONLL-B, the dependence falls away like $m_{c,b}^2/Q^2$ (in the former case the 
whole  ${\cal O}(\alpha_S^2)$ coefficient function is weighted by $m_{c,b}^2/Q^2$, 
while in the FONLL-B scheme the subtraction means that only the massless limit of 
the ${\cal O}(\alpha_S^2)$ coefficient function remains as $m_{c,b}^2/Q^2\to 0$). 
Hence, the dependence on the mass scheme is more limited than in the FFNS at 
NLO, and is particularly small. Indeed, in all but the original TR and TR' schemes,
there is no dependence at high $Q^2$. 

At NNLO the mass scheme dependence
in the FFNS enters in the ${\cal O}(\alpha_S^2)$ and ${\cal O}(\alpha_S^3)$
coefficient functions. In a GM-VFNS it now enters in the 
${\cal O}(\alpha_S^2)$ coefficient functions at low scales, and in boundary 
conditions for evolution, which gives effects which persist to all scales. 
If the GM-VFNS uses the ${\cal O}(\alpha_S^3)$ coefficient functions these will also give 
mass scheme dependent effects at low $Q^2$. However, the expressions for 
the ${\cal O}(\alpha_S^3)$ coefficient functions are themselves still approximations 
\cite{Kawamura:2012cr}. 

Hence, at present it does not seem too important whether the pole mass or 
$\MS$ renormalization scheme is used in a GM-VFNS (indeed in \cite{NNPDF3} 
the pole mass scheme is used, but the $\MS$ values for the masses are taken).
Nevertheless, in future it is probably ideal to settle on the $\MS$ 
mass, since the value of this is quite precisely determined in many experiments, 
which is not true of the pole mass. At the same time it will also be
desirable for different PDF groups to agree on a common value of $m_c$ and 
$m_b$ (which is not the case at present).

\section{Conclusions}

The main purpose of this article is to present and make available PDF sets in the framework used to produce the MMHT2014 PDFs, but with 
differing values of the charm and beauty quark mass. We do not make a 
determination of the optimum values of these masses, but we do investigate and
note the effect the mass variation has on the quality of the fits to the data, 
concentrating on the HERA cross section data with charm or beauty in the final
state. We note that for both the charm and beauty quarks a lower mass
than our default values of $m_c=1.4~\GeV$ and $m_b=4.75~\GeV$ is preferred, 
although these are roughly the values of pole masses one would expect
from conversion from the values measured in the $\MS$ scheme. This 
suggests that in future it may be better to use the $\MS$ definition, though
this is currently not the practice in global fits using a GM-VFNS --- perhaps 
because, as we discuss, the mass scheme dependence has less effect in these 
schemes than for the FFNS. We also make PDFs available with a maximum of 
3 or 4 active quark flavours. The PDF sets obtained for different quark 
masses and for different active quark flavours can be found at~\cite{UCLsite} 
and will be available from the LHAPDF library~\cite{LHAPDF}.

We investigate the variation of the PDFs and the predicted cross sections for standard
processes at the LHC (and Tevatron) corresponding to these variations in 
heavy-quark mass. For reasonable variations of $m_c$ the effects  are small, but not
insignificant, compared to PDF uncertainties. For variations in $m_b$ the
effect is smaller, and largely insignificant, except for the beauty 
distribution itself, which can vary more than its uncertainty at a fixed
value of $m_b$, see, in particular Fig.~\ref{fig:PDFsmbq4}. Hence, currently the uncertainties on PDFs due to quark masses
are not hugely important, but need to be improved in future for very high precision predictions at hadron colliders.

\section*{Acknowledgements}

We particularly thank W. J. Stirling  and G. Watt for numerous discussions on PDFs
and for previous work without which this study would not be possible. We would like
to thank  Mandy Cooper-Sarkar, Albert de Roeck, Stefano Forte, Joey Huston, Pavel Nadolsky and Juan Rojo 
for various discussions on the relation between PDFs and quark masses.  
We would like to thank A. Geiser, A. Gizhko and K. Wichmann for help 
regarding the treatment of uncertainties for ZEUS beauty cross sections.   
This work is supported partly by the London Centre for Terauniverse Studies (LCTS),
using funding from the European Research Council via the Advanced
Investigator Grant 267352. RST would also like to thank the IPPP, Durham, for
the award of a Research Associateship held while most of this
work was performed. We thank the
Science and Technology Facilities Council (STFC) for support via grant
awards ST/J000515/1 and ST/L000377/1.

\bibliography{references}{}
\bibliographystyle{h-physrev}

\end{document}